\documentclass[10pt,final]{siamltex}
\usepackage{amssymb,amsfonts,amsmath,latexsym,dsfont}
\usepackage{tikz,pgflibraryplotmarks}
\usepackage{graphicx}
\usepackage{mathrsfs}
\usepackage[boxed]{algorithm2e}
\usepackage{color}
\usepackage{multirow}
\usepackage{hyperref}
\usepackage{setspace}

\graphicspath{{./}}

\usepackage{boxedminipage}

\usepackage{pgf,pgfarrows} 
\usepackage{subfigure} 

\newcommand{\FrameboxA}[2][]{#2}
\newcommand{\Framebox}[1][]{\FrameboxA}

\newcommand{\mc}[3]{\multicolumn{#1}{#2}{#3}}

\newcommand{\hf}{{\frac 12}}

\newcommand{\bfA}{{\bf A}}

\newcommand{\bfD}{{\bf D}}

\newcommand{\bfH}{{\bf H}}

\newcommand{\bfJ}{{\bf J}}

\newcommand{\bfP}{{\bf P}}

\newcommand{\bfR}{{\bf R}}

\newcommand{\bfe}{{\bf e}}
\newcommand{\bfb}{{\bf b}}

\newcommand{\bfp}{{\bf p}}

\newcommand{\bfz}{ {\bf z}}
\newcommand{\bfv}{ {\bf v}}

\newcommand{\bfu}{{\bf u}}
\newcommand{\bfq}{{\bf q}}

\newcommand{\bfd}{{\bf d}}
\newcommand{\bfm}{{\bf m}}
\newcommand{\bfr}{{\bf r}}

\newcommand{\bfg}{{\bf g}}
\newcommand{\bftau}{{\boldsymbol \tau}}

\newcommand{\bfgamma}{{\boldsymbol \gamma}}

\newcommand{\known}{{\tt\textit{known}}}
\newcommand{\front}{{\tt\textit{front}}}

\renewcommand{\theequation}{\arabic{section}.\arabic{equation}}

\newcommand{\fwi}{{\sf {_{fwi}}}}
\newcommand{\mref}{{\sf {_{ref}}}}
\newcommand{\true}{{\sf {_{true}}}}
\newcommand{\obs}{{\sf {_{obs}}}}
\newcommand{\eik}{{\sf {_{eik}}}}
\newcommand{\joint}{{\sf {_{joint}}}}

\newcommand{\tred}[1]{{#1}}

\begin{document}
\title{Full waveform inversion guided by travel time tomography}



\author{Eran Treister\thanks{Department of Computer Science, Ben-Gurion University of the Negev, Beer Sheva, Israel. ({\tt erant@cs.bgu.ac.il})} \and Eldad Haber\thanks{ Department of Earth and Ocean Sciences, University of British Columbia, Vancouver, Canada. ({\tt haber@math.ubc.ca.})}}
\maketitle

\begin{abstract}
  Full waveform inversion (FWI) is a process in which seismic numerical simulations
  are fit to observed data by changing the wave velocity model of the medium under investigation. The problem is non-linear, and therefore optimization techniques have been used to find a reasonable solution to the problem. The main problem in fitting the data is the lack of low spatial frequencies. This deficiency often leads to a local minimum and to non-plausible solutions. In this work we explore how to obtain low frequency information for FWI. Our approach involves augmenting FWI with travel time tomography, which has low-frequency features. By jointly inverting these two problems we enrich FWI with information that can replace low frequency data. In addition, we use high order regularization, in a preliminary inversion stage, to prevent high frequency features from polluting our model in the initial stages of the reconstruction. This regularization also promotes the non-dominant low-frequency modes that exist in the FWI sensitivity. By applying a joint FWI and travel time inversion we are able to obtain a smooth model than can later be used to recover a good approximation for the true model. A second contribution of this paper involves the acceleration of the main computational bottleneck in FWI--the solution of the Helmholtz equation. We show that the solution time can be reduced by solving the equation for multiple right hand sides using block multigrid preconditioned Krylov methods.
\end{abstract}


\section{Introduction}
Full waveform inversion (FWI) is a process in which the wave propagation velocity of the earth is estimated by using measured wave-field data. The estimation is performed by fitting the field data to simulated data, which are obtained numerically by propagating the wave equation (in time), or by solving the Helmholtz equation (in frequency). The data fitting requires an optimization algorithm, typically a descent algorithm which gradually reduces the misfit between the field and simulated data. To keep the velocity model reasonable, regularization is typically added to the process. However, while the method was investigated over 30 years ago by Tarantola \cite{taran}, it has regained popularity in the last decade when high quality data, computing power and advanced algorithms have been applied to the problem
\cite{pratt1999,EpanomeritakisAkcelikGhattasBielak2008,krebs09ffw,biondi2014simultaneous,van2013mitigating}.
Many algorithms have been proposed for the solution of the problem, however, it remains difficult to solve in practice; solution techniques can be unstable, converging to local minima or to non-plausible solutions.

A naive approach to the FWI problem typically converges to a local minimum. FWI is highly non-linear and is known to be sensitive to its initialization point. As a result,
 most algorithms adopt a frequency continuation strategy. Initially the problem is solved for low frequency data only to obtain a spatially smooth model, then higher and higher frequency data are introduced to the problem to obtain more resolution \cite{pratt1999}. It is known that having low frequency information is crucial if we are to converge to the global minimum. If such information exists in the data, the continuation process is known to be both efficient and stable in producing FWI results. In many realistic scenarios, though, this is not possible because most data acquisition systems still do not collect sufficiently low frequencies in order to effectively use frequency continuation. There may not be sufficient information in the data to lead FWI to a minimizer that geologically makes sense, and for most realistic scenarios FWI converges to a local minimum of the misfit function or to low misfit but non-geological solutions.

The arguments above suggest that in order to find a reasonable minimizer to the misfit without having sufficient prior information for initializing FWI, the physics of the problem needs to be augmented with low frequency information. This can be obtained by solving another inverse problem together with FWI in a joint inversion, complementing the missing low frequencies. In this paper we present this approach using travel time tomography. This problem is computationally attractive when based on the eikonal equation and its factored version \cite{li2013first,benaichouche2015first,TH2016} for forward modeling. Since the solution of the eikonal equation is based on integration of the slowness along rays, the tomographic travel time data contains low frequency information \cite{wang1997sensitivities,virieux2009overview}. Nevertheless, simply using travel time tomography as initialization for FWI may not lead to a feasible recovery \cite{boonyasiriwat2010applications}. In this paper we show that a joint inversion of the two problems, using proper regularization, may lead to a better recovery. In addition, upon successful inversion, the resulting model is faithful to both waveform and travel time data \cite{liu2015joint}. We note that our approach can be combined with other techniques to make the FWI process more robust to its (poor) initial guess, for example tomographic FWI \cite{biondi2014simultaneous} or penalty based constraints relaxation \cite{van2016}.

Using travel time inversion together with FWI alleviates some of the difficulty but may not be sufficient to promote low frequencies at the initial inversion process. To this end, we introduce an adaptive regularization technique that aggressively promotes smoothness initially and is then relaxed in later stages to resolve sharp
features in the model.

Finally, we show how to reduce the computational effort of applying the inversion strategies mentioned above. In this sense, the FWI problem, which involves several frequencies per source, is much more computationally demanding than the travel time tomography. FWI \tred{has} the Helmholtz equation as a forward problem, which is one of the most challenging linear systems to solve. Travel time tomography, on the other hand, requires a rather sophisticated algorithm for treating its forward problem and sensitivities, but is relatively easy to solve computationally. We focus on FWI and compare different strategies for exploiting the large number of sources for reducing the average cost of a linear solve. In particular, we compare block Krylov methods together with multigrid preconditioning on multiple right hand sides.

\bigskip
The rest of the paper is organized as follows. In Section \ref{sec:Inv} we discuss the mathematical formulation, inversion algorithm and forward problem of each of the inverse problems separately. In Section \ref{sec:Joint}, we discuss the joint inversion of the two problems. In Section \ref{sec:MG} we describe how to accelerate the solution of the Helmholtz equation---the main computational bottleneck of FWI, which is the computationally dominant inversion of the two.
Last, in Section \ref{sec:Results}, we demonstrate the use of our method in two and three dimensions,
and show performance results regarding the Helmholtz solution in 3D.
Throughout the paper, scalars and scalar functions are denoted by regular lower case letters. Discretized functions (vectors) are denoted by boldface letters, and matrices are denoted by capital letters.

\section{Mathematical formulation and algorithms}\label{sec:Inv}
In this section we provide the mathematical formulation and numerical algorithms for treating the two inverse problems discussed in this paper: (1) full waveform inversion and (2) travel time tomography. We discuss the mathematical setting and describe the solution and implementation issues of each of the two problems using Gauss-Newton.

In FWI, we typically have sources at many locations and the waveform that is generated by each source is recorded at locations where receivers are placed. These waveform data are usually collected at the receivers over several seconds after the source is generated. FWI can be formulated in either time or frequency domains. In the time domain formulation, the forward problem is the wave equation (assuming constant density) \cite{pratt1998gauss}
\begin{equation}\label{eq:time}
\frac{\partial^2 v(x,t)}{\partial t^2} + \gamma(x)\frac{\partial v(x,t)}{\partial t} - c(x)^2\Delta v(x,t)  = q_s(x,t),
\end{equation}
where $v(x,t)$ is the waveform, $c(x)$ is the wave propagation velocity, $\gamma(x)$ is an attenuation parameter, $\Delta$ is the spatial Laplacian operator. $q_s(x,t)$ is a source function, often formulated as $r(t)q(x)$, where $r(t)$ is the source function in time, and $q(x)$ is a spatial source function. This equation is accompanied by some absorbing boundary conditions \cite{operto20073D}. Forward modeling is performed by applying time-steps over the recorded time \tred{for each source, starting from $v= v_t = 0$ at time, $0$,  before the source is generated}. In the frequency domain formulation of FWI, the observed data undergoes a Fourier transform, \tred{and the forward problem is the Helmholtz equation. Forward modeling is preformed by discretizing the Helmholtz equation as a linear system and solving it for each source and frequency}. \tred{Solving the FWI problem in the frequency domain has several advantages over solving it in the time domain}. First, the size of the time steps in time domain is limited by the Courant–-Friedrichs–-Lewy (CFL) stability condition, \tred{which leads to a large number of time steps} during the forward modeling. As a result, the observed and simulated data are large \tred{and are} hard to handle in memory or communicate over networks. On the other hand, \tred{in frequency-domain, FWI can be} performed by considering data that correspond to only a few frequencies \cite{pratt1999}. Hence, the observed and simulated data are lighter in memory, and it is trivial to separate the forward problem and the inversion to certain frequencies in a continuation process \cite{pratt1998gauss}. In addition, solving the forward problem for multiple sources is inexpensive, assuming a direct solver or an efficient preconditioner can be utilized for solving the resulting linear system. In these cases, the matrix factorization or the preconditioner setup is required only once for all sources. This factorization or setup can be further utilized in the adjoint and Gauss-Newton iteration when considering the inverse problem. Also, iterative solvers can be accelerated using multiple right-hand side techniques \cite{simoncini1995iterative,baker2006improving,el2003block,calandra2012flexible}. The disadvantage of frequency domain formulation lies in the difficulty of solving the Helmholtz equation. While 2D solutions are fairly easy to achieve using direct methods, solving the Helmholtz equation in 3D is considered a numerical challenge, because of the relatively large number of degrees of freedom required to discretize the problem and the indefiniteness of the associated matrix \cite{calandra2013improved,poulson2013parallel,haber2011fast} and its sparsity pattern. We consider the frequency domain FWI for the reasons mentioned above, but generally there is no obvious choice between the two formulations.

\newpage
\subsection{Full waveform inversion in the frequency domain}
To model the waveforms we consider the \emph{discrete} Helmholtz equation as a forward problem, assuming a constant density media
\begin{eqnarray}
\label{eq:helm}
\Delta_{h} \bfu + \omega^{2} (1-i\bfgamma/\omega)\odot\bfm\odot\bfu = \bfq_{s}.
\end{eqnarray}
Here, $\Delta_{h}$ is a discretization of the Laplacian, the symbol $\odot$ represents the Hadamard product of two vectors, $\bfu = \bfu(\bfm,\omega,x_{s})$ is the discrete wavefield, $\bfm$ is the model for the squared slowness (the inverse squared wave velocity $c$ in \eqref{eq:time}), and $\omega$ is the angular frequency, hereafter frequency. The source $\bfq_{s}(\omega)$ is assumed to be the Fourier transform in time of the source function in \eqref{eq:time}. Here it is chosen to be the function $\hat{r}(\omega)\delta(x-x_s)$, where $\hat{r}(\omega)$ is the source wavelet in frequency domain \cite{pratt1999}, and $\delta(x-x_s)$ is a delta function, \tred{which is discretized by the finite volume method}. The vector $\bfgamma>0$ is a parameter used to impose attenuation in the wave propagation model.
The equation is discretized on a finite domain and is accompanied with absorbing boundary conditions that mimic the propagation of a wave in an open domain. To this end, we impose a boundary layer in $\bfgamma$ in addition to attenuation \cite{erlangga2006novel}. In this work we assume that the attenuation parameter is known. 

As noted before, we typically have many sources in FWI, and the waveform that is generated by each source is recorded at the receivers. In our formulation, each observed data corresponding to a source and frequency, is given by
\begin{equation}\label{eq:data}
\bfd^{\fwi\obs}(\omega,x_{s}) = \bfP_s^{\top}\bfu(\bfm_{\true},\omega,x_{s}) + \epsilon
\end{equation}
where $\bfP_{s}$ is a sampling matrix that measures (interpolates) the wave field $\bfu$ at the locations of receivers that record the waveform from sources at $x_{s}$. The data are typically noisy and we assume that the noise, $\epsilon$, \tred{is} Gaussian with zero mean and covariance $\Sigma$.
Given data measurements that are collected for each source and a spectrum of frequencies,
we aim to estimate the true model, $\bfm_{\true}$.
This is done by solving the penalized weighted least squares optimization problem

\begin{eqnarray}\label{eq:minp}
\min_{\bfm_{L} \le \bfm \le \bfm_{H}}&\Phi_{\fwi}(\bfm) = \sum_{j=1}^{n_{f}}{\sum_{i=1}^{n_s}{\left\|\bfP_i^{\top} \bfu_{ij}(\bfm) - \bfd^{\fwi\obs}_{ij}\right\|^2_{\Sigma_{ij}^{-1}}}} + \alpha R(\bfm),
\end{eqnarray}
where $\bfd^{\fwi\obs}_{ij} = \bfd^{\fwi\obs}(\omega_j,x_{i})$ is the observed data that corresponds to the true model in \eqref{eq:data}, and $\bfu_{ij}(\bfm) = \bfu(\bfm,\omega_j,x_i)$ is the waveform for source $i$ and frequency $\omega_j$ that is predicted for a given model $\bfm$, according to the forward problem \eqref{eq:helm}. Recovering the parameters from the data is in many cases \tred{ill-posed. In addition, we have }noise in the data measurements and errors in our numerical modeling. Therefore, we cannot expect to recover the true model, but wish to recover a reasonable estimate of the model by adding prior information \cite{vogel2002computational,somersalo2004statistical}. To accomplish this, we use the bounds $0<\bfm_{L}$ and $0<\bfm_L<\bfm_{H}$, which are forced to keep the model physical, and a regularization term $R(\bfm)$, which is accompanied by a balancing regularization parameter $\alpha$. Because the earth subsurface typically involves layers of different \tred{rock} types, we may assume $\bfm$ to be a layered model, and choose $R$ to promote smooth or piecewise-smooth functions like the total variation regularization term \cite{rudin1992nonlinear}.

To solve the optimization problem \eqref{eq:minp} a variety of methods are typically used. First order methods
such as non-linear conjugate gradient and limited memory BFGS \cite{nw} require little memory, but can converge rather slowly, especially for problems that are highly non-linear where curvature may play a significant role. Our method of choice is the Gauss-Newton method \cite{pratt1998gauss}, which incorporates some curvature
information (we describe this method in the next section). The method converges in fewer steps than the first order methods, but requires repeated solutions of \eqref{eq:helm} with the same model $\bfm$ at each step. Therefore, it is most efficient when an effective solver for \eqref{eq:helm} is at hand, even if it is expensive to construct.

The problem is particularly challenging if the  sources and the receivers
are placed on the same surface and the frequency is high. In this case the problem is known to have
multiple minima and therefore convergence to a local minimum may lead to a model that is very different from the true model.
One way to overcome the problem of local minima is to use continuation techniques, in particular,
frequency continuation which has been shown to be very effective \cite{pratt1998gauss,pratt1999}. The idea is to start the inversion using low frequencies
and to build a smooth background model. The problem is then solved for more frequencies, starting from this background model, and the process continues by adding more frequencies, starting each time from the previous solution. Algorithm \ref{alg:freqCont} summarizes the frequency continuation approach, which seems to be robust and can be initialized with models that are very far from the
final solution. The main problem with this approach is that low frequency data are difficult to measure, making
the approach elegant in principle but difficult to apply in practice.

\begin{algorithm}
\DontPrintSemicolon
\label{alg:freqCont}
\KwSty{Algorithm:}\;
\emph{\# Assume frequencies are given in increasing order: $\omega_1<\omega_2<...<\omega_{n_f}$}\;
Initialize $\bfm^{(0)}$ by some reference model.\;
\For{$k=1,...,n_f$}{
    $\bfm^{(k)} \leftarrow $ Solution of \eqref{eq:minp} using data for $\omega_1,...,\omega_k$, starting from $\bfm^{(k-1)}$.\;
}
\caption{Frequency continuation.}
\end{algorithm}

\subsection{Travel time tomography}

Travel time tomography is a process where the same slowness model $\bfm$ in \eqref{eq:helm} is recovered using first arrival time information.
Assuming that the wave field has a continuous solution of the form $u(x) = a(x) \exp(i \omega \tau(x))$, then by substituting it into the Helmholtz equation \eqref{eq:helm} and assuming that $\omega$ is large, we obtain the eikonal equation
\begin{equation}\label{eq:eik}
|\nabla_{h} \bftau|^{2} - \bfm = 0, \quad \quad \bftau(x_s) = 0,
\end{equation}
where $\bftau$ is the (discrete) first arrival time, $\nabla_{h}$ is a discretization of the gradient operator and $x_s$ is the location of the point source as in \eqref{eq:helm}.
This equation approximately models the first arrivals of the waves at high frequencies.

Similarly to the notation in the previous section, we will refer to the solution $\bftau=\bftau(\bfm,x_s)$ of \eqref{eq:eik} as a function of the source location $x_s$ and the model $\bfm$. The first arrival data  can be extracted from the time domain waveform data, or the high frequency waveform data, either manually or automatically \cite{saragiotis2013automatic}. We now consider the travel time data
\begin{equation}
\bfd^{\eik\obs}(x_{s}) = \bfP _s^{\top}\bftau(\bfm_{\true},x_s) + \epsilon,
\end{equation}
where $\bfP_{s}$ is the same sampling matrix used in \eqref{eq:data}, assuming that both Helmholtz equation and the eikonal equation are discretized on the same mesh. Here $\epsilon$ denotes the noise in the travel time extractions.

Using the new travel time data we can now solve the optimization problem
\begin{eqnarray}\label{eq:mineik}
\min_{\bfm_{L} \le \bfm \le \bfm_{H}}&\Phi_{\eik}(\bfm) = \sum_{i=1}^{n_s}{\left\|\bfP _i^{\top} \bftau_{i}(\bfm) - \bfd^{\eik\obs}_{i}\right\|^2_{\Gamma_{i}^{-1}}} + \alpha R(\bfm),
\end{eqnarray}
where the data term $\bfd^{\eik\obs}_{i} = \bfd^{\eik\obs}(x_i)$, $\bftau_{i}(\bfm) = \bftau(\bfm,x_i)$ and the rest of the parameters (bounds, regularization, noise covariance $\Gamma_i$) are similar to the ones in \eqref{eq:minp}, since the two problems have identical experiment setting and unknown $\bfm$.
\bigskip

There is an important observation to make here.
The eikonal equation does not capture all the physics that is in the wave propagation, and solving it may introduce problems such as caustics and \tred{multiple solutions}. However, our goal here is not to accurately model wave phenomena. The eikonal equation can be
thought of as a reduced model that can capture some of the key features in the media.
Even if some of the data is erroneous due to problems such as multi-pathing, the overall travel time
data are useful
in a least-squares sense to obtain a smooth approximation to the media.

\subsection{Inversion using the projected Gauss-Newton method}
We will now briefly describe the inversion of a problem of the same form as \eqref{eq:minp} and \eqref{eq:mineik} using the Gauss-Newton (GN) method. Assuming that we have $n_{s}$ sources and corresponding observed data $\bfd^{\obs}_{i}$ for each source $i$, let $\bfu_i(\bfm)$ be a predicted field corresponding to a model $\bfm$, according to a forward modeling equation (either Helmohotz or eikonal), and let the
matrix $\bfP^{\top}_i$ be a projection matrix that injects the field $\bfu_i$ from the whole space to the measurement points.
The regularized data fitting problem is given by
\begin{eqnarray}
\label{eq:optinv}
\min_{\bfm}\ \Phi(\bfm) = \hf \sum_{i=1}^{n_{s}} \| \bfP_i^{\top} \bfu_{i}(\bfm) - \bfd_{i}^{\obs} \|^{2}_{\Sigma_i^{-1}} + \alpha R(\bfm),
\end{eqnarray}
with regularization parameters similar to \eqref{eq:minp}.

At each iteration $k$ of GN we obtain the approximation
\begin{equation}\label{eq:GNapprox}
\bfu_i(\bfm^{(k)}+\delta\bfm) \approx \bfu_i(\bfm^{(k)}) + \bfJ_i(\bfm^{(k)})\delta\bfm,
\end{equation}
where $\bfJ_i = \nabla_{\bfm} \bfu_i$ is the Jacobian matrix, also referred to as  the sensitivity.
It is well known \cite{hao2000} that $\bfJ_{i}$ need not be formed or stored and only matrix vector products
of $\bfJ_{i}$ and its transpose are calculated by solving the forward and adjoint problems.

At each step, we place \eqref{eq:GNapprox} in \eqref{eq:optinv} and get an alternative minimization
\begin{eqnarray}\label{eq:GNminp}
\min_{\delta\bfm}\ \hf \sum_{i=1}^{n_{s}} \| \bfP_i^{\top} \left(\bfu_i(\bfm^{(k)}) + \bfJ_i(\bfm^{(k)})\delta\bfm\right) - \bfd_{i}^{\sf obs} \|^{2} + \alpha R(\bfm^{(k)}+\delta\bfm).
\end{eqnarray}
To solve it, we essentially compute the gradient
\begin{eqnarray}\label{eq:gradpi}
\nabla_{\bfm}\Phi(\bfm^{(k)}) = \sum_{i=1}^{n_{s}} \bfJ_i(\bfm^{(k)})^{\top}\bfP_i( \bfP_i^{\top} \bfu_{i}(\bfm^{(k)}) - \bfd_{i}^{\obs}) + \alpha \nabla_{\bfm} R (\bfm^{(k)}),
\end{eqnarray}
and to get the step $\delta\bfm$ we approximately solve the linear system
\begin{eqnarray}\label{eq:Hessian}
\nonumber& \bfH \delta \bfm = -\nabla_{\bfm}\Phi(\bfm^{(k)}), \\
&\bfH =  \sum_{i=1}^{n} \bfJ_i(\bfm^{(k)})^{\top} \bfP \bfP^{\top} \bfJ_i(\bfm^{(k)}) + \alpha \nabla^{2} R(\bfm^{(k)}).
\end{eqnarray}

The linear system is solved using the Preconditioned Conjugate Gradient (PCG) method where only matrix vector products are required to obtain the solution.
Finally, the model is updated, $\bfm \leftarrow \bfm + \mu \delta \bfm$ where $\mu \le 1$ is a line search parameter that is chosen so that the objective function is decreased at each iteration. \tred{We note that the terms $\nabla_{\bfm}R$ and $\nabla_\bfm^{2}R$ only exist for smooth regularization functions, but these can be approximated for many popular regularization terms like Total Variation; see \cite{haber2014computational} for more details}.

To treat a bound-constrained version of \eqref{eq:optinv}, which is similar to the optimization problems \eqref{eq:minp} and \eqref{eq:mineik}, we use the projected GN method, which is also described in \cite{haber2014computational}. At each iteration, the update $\delta\bfm$ is divided into active and inactive sets. On the inactive set, $\delta\bfm$ is defined by approximately solving \eqref{eq:GNminp} by the projected PCG method. On the active set, the update $\delta\bfm$ is computed by projected steepest descent (see \cite{nw} for more details.)

\subsection{The Helmholtz forward problem and its sensitivities} As mentioned before, in the frequency domain version of FWI, we use a discretized version of the forward problem \eqref{eq:helm} for forward modelling. We use a second-order finite difference scheme on a regular grid, resulting in a sparse linear system
\begin{equation}\label{eq:LinearSystem}
A(\bfm,\omega)\bfu=\bfq.
 \end{equation}
We are mostly focused on problems where the frequency $\omega$ is high. In this case, the resulting linear system is highly indefinite and difficult to solve. We note that using this discretization, one has to respect the rule of having at least ten grid-point per wavelength \cite{erlangga2006novel,haber2011fast,calandra2013improved,poulson2013parallel}, otherwise the numerical solution is polluted by phase errors. To solve \tred{the linear system \eqref{eq:LinearSystem}} we either use direct solvers like \cite{MUMPS2001,schenk2004solving} or iterative methods like \cite{erlangga2006novel,oosterlee2010shifted,cools2014new,haber2011fast}. If the problem is two-dimensional, the mentioned direct solvers are preferred, but if the problem is three-dimensional and large, we use a variant of the shifted Laplacian multigrid approach.
We summarize this approach later in Section \ref{sec:MG}, and discuss some techniques to reduce the computational effort of solving this equation for multiple right hand sides.

\bigskip
The sensitivity matrix for \eqref{eq:helm}, required in \eqref{eq:gradpi}-\eqref{eq:Hessian}, is given by
\begin{eqnarray}
\label{eq:fwisens}
\bfJ\fwi(\bfm) =   -\omega^2 A(\bfm,\omega)^{-1} \diag(1-i\bfgamma/\omega)\, \diag(A(\bfm,\omega)^{-1}\bfq).
\end{eqnarray}
This is a dense matrix, which cannot be stored in memory, but it can be implicitly applied to a vector. If one has the fields $\bfu_{ij} = A(\bfm,\omega_j)^{-1}\bfq_{s_{i}}$, stored in memory, then multiplying $\bfJ\fwi$ with a vector can be obtained using one linear system solve for each pair of source and frequency. If the fields $\bfu_{ij}$ are too memory consuming, we save them to the disk and apply the sensitivity in batches of sources. The fields can be stored in low precision either in the memory or on the disk (we found a 16-bit floating point representation to be sufficient). Because the fields are complex this results in a storage of 32-bit per grid unknown per source-frequency pair.
Of course, when solving the Helmholtz equation with an iterative solver, the linear systems should not be solved to machine precision, but some error tolerance can be allowed---see \cite{van20143d} for a discussion about the choice of these error tolerances.

\subsection{The eikonal forward problem and its sensitivities} To numerically solve the eikonal equation in \eqref{eq:eik}, we use its factorized version, which is known to yield more accurate solutions for point sources \cite{fomel2009fast,luo2011factored,luo2012fast,LouQianBurridge2014}. That is, we set $\bftau = \bftau_0\bftau_1$ in \eqref{eq:eik}, where $\bftau_0$ is a discretization of the function $\tau_0 = \|x-x_s\|_2$ is the distance function from the source---this is the solution of \eqref{eq:eik} for $\bfm=1$. We then solve the following equation for $\bftau_1$
\begin{equation}\label{eq:factoredeikonal}
|\bftau_0\odot \nabla_{h}\bftau_1 + \bftau_1\odot \bfp_0|^2 = \bfm,
\end{equation}
where $\bfp_0$ is the discretized $\nabla\tau_0$ which is obtained analytically. Following \cite{TH2016}, we discretize the factored equation using the following Gudonov upwind scheme (here we show the discretization in 2D and the discretization in 3D is a straightforward extension):
\begin{equation}\label{eq:Gudonov}
 \left[\max\{\hat{\bfD}^{-x}_{ij}\bftau_1,-\hat{\bfD}^{+x}_{ij}\bftau_1,0\}^2 + \max\{\hat{\bfD}^{-y}_{ij}\bftau_1,-\hat{\bfD}^{+y}_{ij}\bftau_1,0\}^2\right]= m_{ij},
\end{equation}
where the matrices $\hat{\bfD}$ are the discretized factorized finite difference derivative operators. For example, the backward first order factored derivative operator is given by
\begin{equation}\label{eq:FOfactoredD}
\hat{\bfD}^{-x}_{ij}\bftau_1 = (\tau_0)_{ij}\frac{(\tau_1)_{i,j}-(\tau_1)_{i-1,j}}{h} + (p^x_0)_{ij}(\tau_1)_{ij},
\end{equation}
where $\bftau_0$ and $\bfp^x_0 = \frac{\partial\tau_0}{\partial x}$ are known (see \cite{TH2016}
for details).

To solve this nonlinear partial differential equation efficiently we use the Fast Marching (FM) method in \cite{TH2016}. This method is based on the FM method in \cite{sethian1996fast,sethian1999fast}, which was suggested for the non-factored eikonal equation. All of these methods use the monotonicity of the solution along the characteristics in order to obtain an efficient nonlinear Gauss-Seidel solver. This method uses an upwind scheme and start from the source, updating the values of $\bftau_1$ everywhere on the grid. It uses two sets, \front and \known, where \known are the already computed ``known'' values and \front is the set of points at the front of the wave propagation. At each iteration, FM chooses the point with the minimal value of $\bftau$ in \front and moves it to \known. Then FM updates the values of the neighbors of this point according to \eqref{eq:Gudonov} and puts them in \front. Finding the minimal value of \front is obtained using minimum heap. Algorithm \ref{alg:FM} summarizes the FM method.

\begin{algorithm}
\DontPrintSemicolon
\label{alg:FM}
\KwSty{Algorithm: Fast Marching }\;
Initialize:
$\tau_{ij} = \infty$ for all $ij$, $\bftau(x_s) = 0$,
\known$\leftarrow\emptyset$, \front$\leftarrow\{x_s\}$.\;
\While{\front$\neq\emptyset$}{
    Find $x_{i_{min},j_{min}}$: the minimal entry in \front:\;
    Add $x_{i_{min},j_{min}}$ to \known and take it out of \front.\;
    Add the unknown neighborhood of $x_{i_{min},j_{min}}$ to \front.\;
    \textbf{for each} unknown neighbor $x_{i,j}$ of $x_{i_{min},j_{min}}$\;
    $\quad$Update $\tau_{ij}$ by solving Eq. \eqref{eq:Gudonov}, using only entries in \known. \;
    \textbf{end}
}
\end{algorithm}

\bigskip

To obtain the sensitivity we first rewrite \eqref{eq:Gudonov} using the same derivative operators (forward or backward) that are chosen by the FM algorithm for each grid point $(i,j)$ when solving the forward problem for $\bfm$. In the points where no derivative is chosen in the solution of \eqref{eq:Gudonov}, a zero row is set in the corresponding operator. Once the operators $\hat \bfD$ are set by the choices of FM, the sensitivity is given by
\begin{eqnarray}\label{eq:Sensitivity}
\bfJ\eik(\bfm,\bftau_1) = (\diag(2\hat{\bfD}^{x}\bftau_1)\hat{\bfD}^{x} + \diag(2\hat{\bfD}^{y}\bftau_1)\hat{\bfD}^{y})^{-1}.
\end{eqnarray}
Here, $\hat{\bfD}^{x}$ and $\hat{\bfD}^{y}$ are the matrices that apply the finite difference derivatives in \eqref{eq:Gudonov} according to the choice of FM.

The matrix \eqref{eq:Sensitivity} is defined by an inverse of a sparse matrix that involves only derivatives in $x$ and $y$ directions. Multiplying the sensitivity matrix with an arbitrary vector $\bfv$ (computing $\bfz = \bfJ\eik \bfv$) is equivalent to solving the (sparse) linear system
\begin{equation}\label{eq:Sensitivity2}
 (\diag(2\hat{\bfD}^{x}\bftau_1)\hat{\bfD}^{x} + \diag(2\hat{\bfD}^{y}\bftau_1)\hat{\bfD}^{y}) \bfz  =\bfv.
 \end{equation}
Since the FM algorithm uses only known variables for determining each new variable,
by reordering the unknowns according to the FM order we obtain a lower triangular system
which can be solved  efficiently in one forward substitution sweep in $O(n)$ operations. For more information on the solution of travel time tomography using FM see \cite{li2013first,TH2016}.

Applying the operator \eqref{eq:Sensitivity} is required for each source in \eqref{eq:mineik}, but storing storing the sparse matrix in \eqref{eq:Sensitivity2} for each source may be highly memory consuming at large scales. Therefore, we only save the values that are necessary for applying the matrix and solve the lower linear system by calculating each row when needed. This requires the travel time factor $\bftau_1$, the ordering of FM, and the direction of each forward/backward derivative. Based on experience, we have found that holding $\bftau_1$ in single precision is sufficient. Assuming the grid is not too large, so that the number of unknowns is smaller than $2^{32}$, we hold the ordering of variables using 32-bit unsigned integer. We hold the direction in a 8-bit integer, that is able to support the $5^3$ options for the finite-difference directions in 3D. This results in a 72-bit memory per grid unknown per source for applying the sensitivities.

\section{Joint FWI and travel time tomography}
\label{sec:Joint}

As mentioned before, it is common to solve the FWI problem \eqref{eq:minp} by the frequency continuation process in Algorithm \ref{alg:freqCont}. However, in the absence of low frequency data this process often converge to a local minimum. Our approach to overcoming the lack of low-frequency data is to jointly invert \eqref{eq:minp} with a complementary problem like the travel time tomography problem in Eq. \eqref{eq:mineik}. The joint problem is formulated by
\begin{eqnarray}\label{eq:minjoint}
\nonumber
\min_{\bfm_{L} \le \bfm \le \bfm_{H}}\Phi_{\joint}(\bfm) &=&  \sum_{j=1}^{n_{f}}{\sum_{i=1}^{n_s}{\left\|\bfP_i^{\top} \bfu_{ij}(\bfm) - \bfd^{\fwi\obs}_{ij}\right\|^2_{\Sigma_{ij}^{-1}}}}
  \\& +&
 \beta\sum_{i=1}^{n_s}{\left\|\bfP_i^{\top} \bftau_{i}(\bfm) - \bfd^{\eik\obs}_{i}\right\|^2_{\Gamma_{i}^{-1}}}
  + \alpha R(\bfm)
\end{eqnarray}
where all the components are exactly as in \eqref{eq:minp} and \eqref{eq:mineik}, and $\beta>0$ is the balancing parameter between the two problems.

There are two advantages to the optimization problem in \eqref{eq:minjoint} compared to \eqref{eq:minp}. First, assuming that the data does not contain low frequencies, travel time can be substituted for low frequency waveform data.
Second, the resulting model honors {\em both} the travel time equations
and the full waveform equations, thus, so it is consistent with different physical interpretations of
the wavefield. Computationally, the cost of adding the tomography to \eqref{eq:minp} and obtaining its
sensitivities is small compared to the corresponding operations for the Helmholtz equation (\ref{eq:helm}), so the additional computational cost in solving (\ref{eq:minjoint}) instead of (\ref{eq:minp}) is negligible.

The problem is non-convex and typically has many local minima, therefore, when
solving the optimization problem we use a process that is akin to the frequency continuation
previously discussed, replacing the missing low frequency data with travel time data.
If the sources and receivers are all on the surface, then
the first arrival information typically depends only on the top part of the medium.
If one applies the travel time tomography on its own, the bottom part of the model may change due only to regularization, without any relation to the true model or the data. Correcting this later in the continuation process with high frequency FWI may not be possible, which is one of the reasons initializing FWI with a travel time tomogram may result in a local minimum
\cite{boonyasiriwat2010applications}. For this reason, we start the continuation by solving
\eqref{eq:minjoint} for {\em both} travel time as well as the lowest available frequency, using a proportional weight $\beta\gg 1$. This way, the obtained model satisfies part of the waveform data so that the gap is not too high when introducing more frequencies. Still, if the first available frequency is too high (which is often the case), this still may result in poor recovery. To obtain a relatively robust process we must be able to extract some low-frequency information from FWI alone, so that the recovery of a smooth model is not guided only by the tomography. We achieve this by using high-order regularization (together with the joint inversion), which is discussed in the next section. We note that the frequency gap problem may be further relieved by adding processes that are more sophisticated than travel time tomography, like stereotomography \cite{lambare2004stereotomography,lambare2008stereotomography}, but it is generally difficult to predict when this ``frequency gap'' will closed.

\subsection{High-order regularization}

We now examine the difficulty of recovering a true model $\bfm_{\true}$ without having low frequency data.
Let the forward problem in frequency domain be $F(\bfm,\omega)$, and the measured data be $\bfd^{\obs}(\omega)$. Assume that we measure the distance between the computed data and the observed data by a misfit functional $S(\bfd(\bfm,\omega),\bfd^{\obs}(\omega))$. For example, $S$ can be a least square function as in \eqref{eq:minp}, or a more complicated one that is based on phase shift \cite{virieux2009overview}, Wiener transform \cite{WarnerEtAt2013}, or optimal mass transport \cite{metivier2016measuring}. FWI algorithms are designed to minimize $S$ with additional regularization. Now assume that we are given some initial reference model, $\bfm_0$, such that the error $\bfm_{\true}-\bfm_0$ \tred{contains low wavenumber components}. Any gradient-based method uses the gradient of the misfit $S$ with respect to the model in order to compute a step from $\bfm_0$ (either directly or scaled by an approximation of the Hessian inverse).
The gradient is given by $\bfJ^{T}\partial_{\bfd} S$, where $\bfJ$ is the sensitivity matrix, $\bfJ = \partial_{\bfm} F(\bfm)$. In cases where the singular vectors of $\bfJ$ that correspond to low spatial frequencies have small singular values, the gradient itself also contains little of these low frequencies, independent of the choice of $S$, so that any gradient-based correction to $\bfm_0$ cannot reduce the low-frequency component of the error.

Let us look now specifically at the problem \eqref{eq:minp} (and its joint version \eqref{eq:minjoint}). Denote the Helmholtz operator in \eqref{eq:helm} by $A$ (ignoring boundary conditions), i.e., $A = \Delta_h + \omega^{2} \mbox{diag}{(\bfm)}$, discretized by a second order finite difference scheme.
\tred{The Jacobian $\bfJ$ in \eqref{eq:fwisens} has the inverse of $A$ multiplied by a diagonal matrix that contains the wave field $\bfu$. This is, it is the inverse of the Helmholtz operator multiplied by an oscillatory diagonal scaling. Even if we ignore this diagonal scaling, $\bfJ$ has dominating singular vectors which have oscillatory spatial behavior that correspond to a frequency $\omega$. The magnitude of these components is roughly the magnitude of the reciprocal of the associated eigenvalue of $A$. Consider a simple Fourier component $\exp{(i\frac{\theta x}{h})}$ in 1D. Assuming a constant medium $\bfm=1$, the eigenvalues of $A$ are $(2\cos\theta - 2 + \omega^2)$, for $\theta\in[-\pi,\pi]$. The value of $\theta$, for which this eigenvalue expression is close to zero, dictates the spatial oscillatory behavior of the dominating singular vector of $\bfJ$. Consequently, the largest eigenvalues of $\bfH = \bfJ^\top\bfJ$ are approximately the square of the leading eigenvalues of $A^{-1}$
that are highly oscillatory (again ignoring the oscillatory diagonal scaling). Because we solve the GN problem \eqref{eq:Hessian} using a few iterations of a Krylov method (CG), the solution $\delta\bfm$ is generally comprised by the dominating eigenvectors of the matrix $\bfH$, which is expected to lead to an oscillatory $\delta\bfm$ (recall that $\delta\bfm\in\mbox{span}\{\bfg,\bfH\bfg,\bfH^2\bfg,\bfH^3\bfg,...\}$, where $\bfg$ is the right hand side of \eqref{eq:Hessian}).}

To generate an initial smooth model, we propose a preliminary stage where we solve \eqref{eq:minjoint} using the high order regularization
\begin{equation}\label{eq:reg1}
R_{1}(\bfm) = \|\Delta_h(\bfm-\bfm_{\mref})\|_2^2.
\end{equation}
This type of regularization is also known as spline smoothing (see \cite{WahbaBook}). Following \cite{haber2014computational}, we precondition the Hessian by $\nabla_\bfm^2R(\bfm)$ during the PCG iterations for solving \eqref{eq:Hessian} to prevent oscillatory components from entering the model. The Fourier symbol of the preconditioned Hessian is approximately
$$((2-2\cos\theta)(2\cos\theta - 2 + \omega^2))^{-2},$$
for which the smooth spatial mode that corresponds to $\theta\approx0$ is significant as the oscillatory modes. We use this for the waveform inversion of the first one or two frequencies (also with the travel time). After we obtain a smooth initial guess, we can continue to invert the joint problem using frequency continuation and a standard regularization
\begin{equation}\label{eq:reg2}
R_{2}(\bfm) = \|\nabla_h(\bfm-\bfm_{\mref})\|_2^2,
\end{equation}
 allowing oscillatory modes. Alternatively, one can use total variation at this stage, which further allows for the recovery of piecewise smooth models. A complete description of our two-stage inversion algorithm is given in Algorithm \ref{alg:twoPhaseJointInv}.
\begin{algorithm}[tb]
    \label{alg:twoPhaseJointInv}
   \caption{Two stage joint inversion. }
   \DontPrintSemicolon
   \KwSty{\textit{Stage I: continuation for a smooth initial model.}}\;
   \For{$f = 1,2,...,f_{low}$ }{
    Solve problem \eqref{eq:minjoint} for frequencies $\omega_1$ to $\omega_{f}$ using \eqref{eq:reg1}\;
   }
   \KwSty{\textit{Stage II: continuation for final inversion.}}\;
   \For {$f = 1,2,..,$ all frequencies}{
    Solve problem \eqref{eq:minjoint} for frequencies $\omega_1$ to $\omega_{f}$ using \eqref{eq:reg2}\;
   }
\end{algorithm}

\section{Solving the Helmholtz equation for multiple right hand sides}
\label{sec:MG}
In this section we describe the multigrid algorithm we use for solving \eqref{eq:helm}. Generally a multigrid approach aims at solving a linear system $\bfA\bfu=\bfq$ iteratively by using two complementary processes: relaxation and coarse-grid-correction. The relaxation is usually obtained by a standard iterative method like Jacobi or Gauss-Seidel and is only effective for reducing certain components of the error. In particular, such relaxation is effective at reducing the error spanned by the eigenvectors of $\bfA$ that correspond to relatively high eigenvalues. To treat the entire spectrum, multigrid methods also use a coarse grid correction, where given some iterate $\bfu^{(k)}$, the linear system for the corresponding error and residual $\bfA\bfe = \bfr = \bfq - \bfA\bfu^{(k)}$ is projected onto the subspace of a prolongation matrix $\hat{\bfP}$ and a restriction matrix $\hat{\bfR}$ = $\hat\bfP^T$. This results in a coarser linear system $\bfA_c\bfe_c = \bfr_c$, where $\bfA_{c}$ is the differential operator on a course grid, and $\bfr_c = \hat{\bfP}^T\bfr$ (the subscript $c$ denotes coarse components). In this work we define $\bfA_{c} = \hat\bfP^T\bfA\hat\bfP$. Algorithm \ref{alg:TwoCycle} summarizes the process. By treating the coarse problem recursively, we obtain the multigrid V-cycle, and by treating the coarse problem recursively twice (by two recursive calls to V-cycle) we obtain a W-cycle. For more information see \cite{BHM00,TOS01,Yav06} and references therein.

\begin{algorithm}
\DontPrintSemicolon
\label{alg:TwoCycle}
\KwSty{Algorithm: Two-grid Cycle}\;
\begin{enumerate}\Indm\Indm
\item Apply pre-relaxations: $\bfu \leftarrow Relax(\bfA,\bfu,\bfq)$\;
\item Define and restrict the residual $\bfr_c = \hat{\bfP}^T(\bfb - \bfA\bfu)$.
\item Define $\bfe_c$ as the solution of the coarse-grid problem $\bfA_c\bfe_c=\bfr_c$.
\item Prolong $\bfe_c$ and apply coarse grid correction: $\bfu \leftarrow \bfu + \hat{\bfP}\bfe_c$.
\item Apply post-relaxations: $\bfu \leftarrow Relax(\bfA,\bfu,\bfb)$.
\end{enumerate}
\caption{Two-grid cycle.}
\end{algorithm}

To treat the Helmholtz equation \eqref{eq:helm} using shifted Laplacian multigrid, one introduces a complex shift to the system by adding a positive constant to $\bfgamma$ in \eqref{eq:helm}. From a physical point of view, this damps the amplitude of the wave, and from an algebraic point of view, this reduces the condition number of the matrix. The shifted Laplacian approach involves applying the shifted matrix as a preconditioner for the original system \eqref{eq:helm} inside a Krylov method. The preconditioning is obtained by applying a multigrid cycle for inverting the shifted matrix. The Krylov methods of choice are usually BiCGSTAB \cite{van1992bi} or (flexible) GMRES \cite{saad1993flexible}. In this work the prolongation and restriction operators are defined as bilinear interpolation and full-weighting operators respectively.

The shifted Laplacian multigrid method is very popular and some software approaches have been proposed for implementing it, e.g., \cite{knibbe2011gpu,Tobias2016}. Here we discuss how to improve its performance by exploiting the multiple right hand sides that are included in FWI. That is, given a large number of right hand sides, we aim to reduce the \emph{average} solution time of \eqref{eq:helm} per right hand side. We use a large multicore machine with a large RAM---a configuration that is relatively common. Because the right-hand-sides are many, we give less emphasis to the setup time and focus on the solution time given the multigrid hierarchy.

The computational ingredients involved in the multigrid version of Algorithm \ref{alg:TwoCycle} as a preconditioner to a Krylov method include sparse matrix-vector multiplication, vector operations (addition, substraction and inner-products) and the coarsest grid solution. Each of the ingredients above has to be accelerated using a shared memory multicore computation. To this end, we consider the block versions of Krylov methods mentioned above \cite{simoncini1995iterative,baker2006improving,el2003block,calandra2012flexible}. Working with such methods enables us to exploit level-3 BLAS parallelism for the vector operations, and improve the sparse-matrix products when applied to multiple vectors. To precondition a block of right hand sides, we apply a block multigrid cycle where all the relaxations, restrictions and interpolations are performed on blocks. We note that more effort is involved in applying the block Krylov methods as the size of the blocks grow. On the other hand, the convergence properties of the solver improve as the block size increases.

Unlike most multigrid scenarios, because of the sign-changing nature of the smooth error modes of the Helmholtz operator at high frequency, we can coarsen the system only a few times using standard transfer operators. At large scales, the coarsest grid operator is of considerable size and the cost of the solution of the coarsest grid is high. This is accelerated by applying the LU solution for a block of right hand sides. Generally, the larger the blocks, the better the parallelism performs, and with less overhead because of changes between the different parallel codes. Lastly, we note that in order to solve the transposed system in \eqref{eq:fwisens}, we may use multigrid preconditioning using a transposed hierarchy without recomputing the hierarchy between solving the original system and its transpose.

\section{Numerical experiments}
\label{sec:Results}

In this section we demonstrate the joint FWI and travel time tomography inversion, in two and three dimensions using the SEG/EAGE salt model \cite{aminzadeh19973}. 
All of our experiments were conducted on a single workstation operating Ubuntu 14.04 with 2 $\times$ Intel Xeon E5-2670 v3 2.3 GHz CPUs using 12 cores each, and a total of 128 GB of RAM. Our code is written in Julia compiled with Intel Math Kernel Library (MKL). The only external packages that we use are MUMPS, and \href{https://github.com/JuliaInv/ParSpMatVec.jl}{\texttt{ParSpMatVec.jl}} a parallel sparse matrix-vector multiplication written in Fortran. We use the inversion package jInv.jl \cite{jInv16}, which is freely available from its Github page (\url{https://github.com/JuliaInv/jInv.jl}). Our FWI, travel time tomography and multigrid packages, which are add-ons to jInv, are available online at \href{https://github.com/JuliaInv/FWI.jl}{\texttt{FWI.jl}}, \href{https://github.com/JuliaInv/EikonalInv.jl}{\texttt{EikonalInv.jl}} and \href{https://github.com/JuliaInv/Multigrid.jl.git}{\texttt{Multigrid.jl}}.

\subsection{Joint inversion in two dimensions}

\begin{figure}
\begin{center}
	\newcommand{\image}[1]{\includegraphics[width=0.48\linewidth]{#1}}
  \subfigure[\footnotesize The true 2D SEG/EAGE velocity model.]{\image{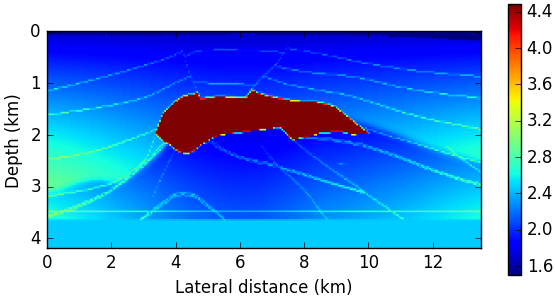}\label{fig:2D-1}}
  \subfigure[\footnotesize The starting linear model.]{\image{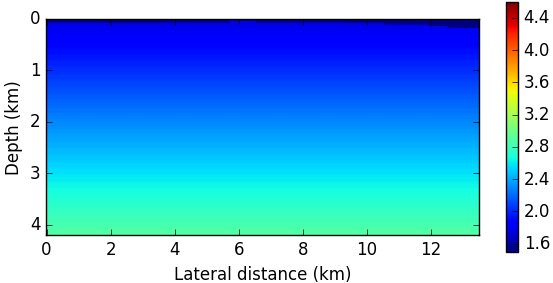}\label{fig:2D-2}}
\end{center}
\caption{\footnotesize The two dimensional SEG/EAGE salt body model and the linear reference model. Velocity units are in $km/sec$.}
\label{fig:SEG_set}
\end{figure}

For our first experiment, we use a 2D slice of the SEG/EAGE model, presented in Figure \ref{fig:2D-1}, using a $600\times300$ grid representing an area of approximately 13.5 km $\times$ 4.2 km. We wish to recover this model using only the linear model in Figure \ref{fig:2D-2} as an a priori reference. Because we use an artificial layer to prevent reflections from the domain boundaries we add a padding of 30 grid points to populate most of the layer in each boundary of the model (except the top free surface)---resulting in a $660\times330$ grid.

\begin{figure}
\begin{center}
	\newcommand{\image}[1]{\includegraphics[width=0.48\linewidth]{#1}}
    \image{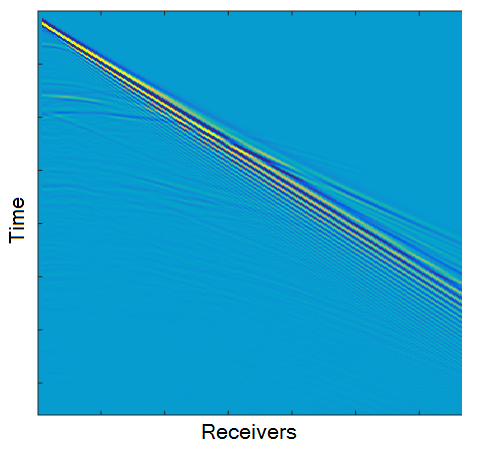}
\end{center}
\caption{\footnotesize A snapshot of the first 8 seconds $\times$ 8 kilometers of the generated time-domain data.}
\label{fig:time}
\end{figure}

\begin{figure}
\begin{center}
	\newcommand{\image}[1]{\includegraphics[width=0.96\linewidth]{#1}}
  \image{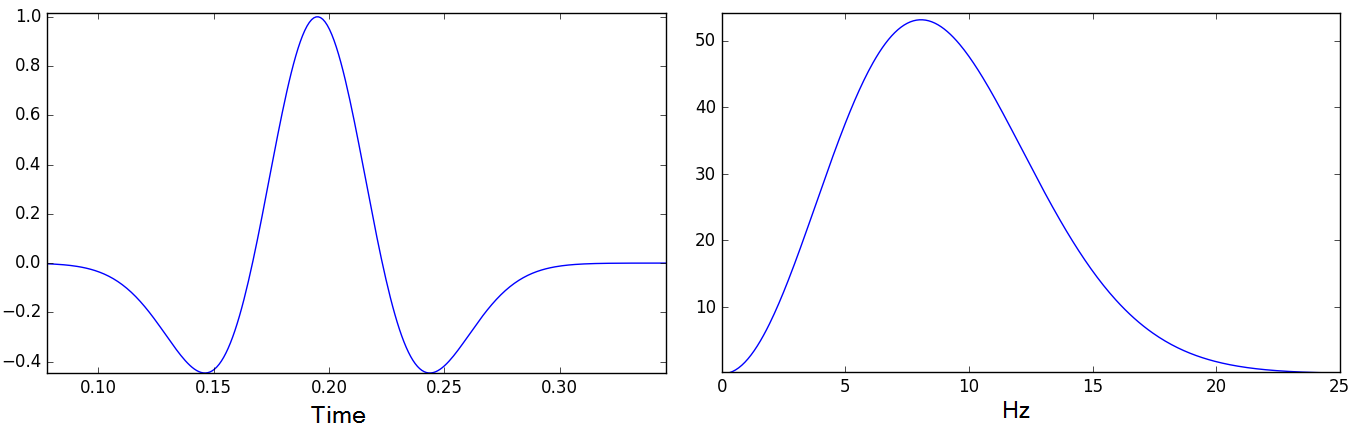}
\end{center}
\caption{\footnotesize The Ricker wavelet source function in time (left) and frequency (right) domains.}
\label{fig:ricker}
\end{figure}

\begin{figure}
\begin{center}
	\newcommand{\image}[1]{\includegraphics[width=0.45\linewidth]{#1}}
    \subfigure[\footnotesize A wave field for $f=2.0$ Hz]{\image{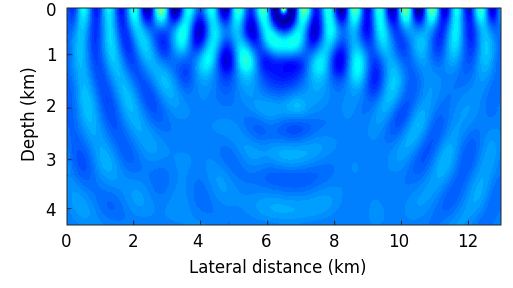}\label{fig:low}}
    \subfigure[\footnotesize A wave field for $f=7.0$ Hz]{\image{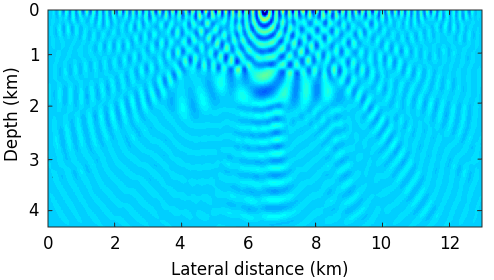}\label{fig:high}}
\end{center}
\caption{\footnotesize Examples of frequency domain wave fields obtained for the SEG/EAGE model. $x$ and $y$ axes denote the number of grid points that we use, including the padding.}
\label{fig:fields}
\end{figure}

\begin{figure}
\begin{center}
	\newcommand{\image}[1]{\includegraphics[width=0.90\linewidth]{#1}}
    \image{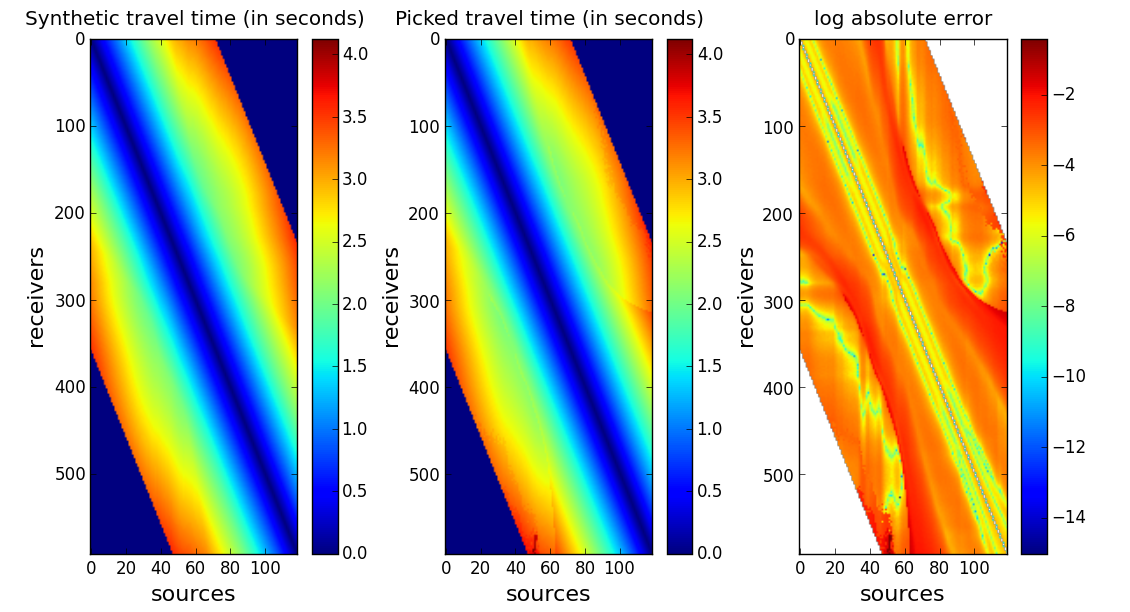}
\end{center}
\caption{\footnotesize The synthetic travel time data obtained by the forward eikonal solver versus the automatically picked travel time. The log absolute error appears on the rightmost image.}
\label{fig:travelPick}
\end{figure}

In this experiment we generate our frequency domain and travel time data from synthetic time-domain waveform data for an array of 119 equally spaced sources and 592 equally spaced receivers located on the top row of the model. The time domain data is generated by discretizing \eqref{eq:time} using second order central difference schemes, and performing time steps using the leapfrog scheme from $t=0$ to $18$ seconds with time steps of 1 $ms$ per step. For the source function $r$, we use the Ricker wavelet function \cite{ricker}.
$$
r(t) = (1-2\pi^2f_M^2t^2)\exp(-\pi^2f_M^2t^2),
$$
centered around frequency $f_M=8Hz$. The Ricker function and its Fourier transform $\hat{r}(\omega)$ appear in Figure \ref{fig:ricker}. For $\bfgamma$, we use an attenuation of $0.01\cdot4\pi$, and use 34 grid points for the absorbing layer itself (30 of which lie in the model padding). Figure \ref{fig:time} shows a snapshot of the generated time domain data.
To this trace we added a Gaussian white noise of the strength of $1\%$ of the data, and applied a Fourier transform to get the frequency domain data.
We extracted frequency domain wave fields corresponding for frequencies $f_i = \{2.0,2.5,3.5,4.5,6.0\}$ $Hz$ ($\omega_i = 2\pi f_i$). In addition, we picked the travel times from the traces in time domain by using an automatic process adapted from the Modified Coppens’s method \cite{sabbione2010automatic}. Figures \ref{fig:low} and \ref{fig:high} show wave fields that correspond to one of the sources for frequencies $f_i$ 2.0 and 7.0 respectively. The data for each source are recorded only in receivers distanced 50 m to 8 km away on the free surface. Figure \ref{fig:travelPick} shows the travel time data and the errors in the automatic time picking compared to a synthetic travel time data.

\begin{figure}
\begin{center}
	\newcommand{\image}[1]{\includegraphics[width=0.48\linewidth]{#1}}
    \subfigure[\footnotesize FWI reconstruction]{\image{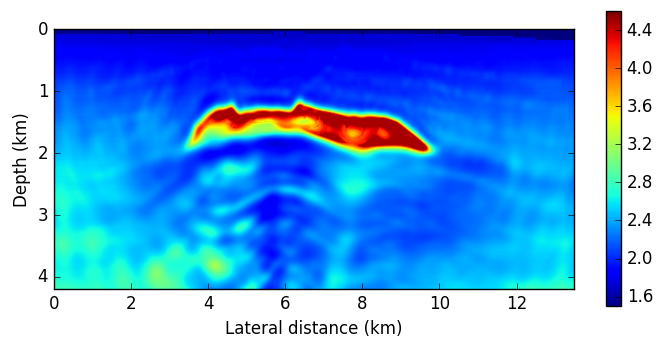}\label{fig:mFWIonly}}\\
  \subfigure[\footnotesize Travel time reconstruction]{\image{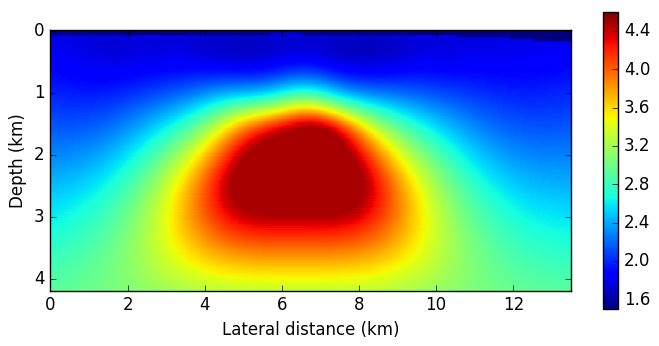}\label{fig:mTomoI}}
    \subfigure[\footnotesize The final model obtained by FWI initialized with the travel time reconstruction.]{\image{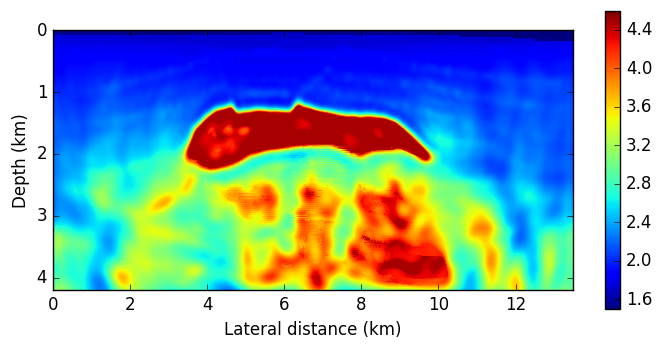}\label{fig:mfwiTomoII}}
    \subfigure[\footnotesize The model obtained in stage I of the joint inversion.]{\image{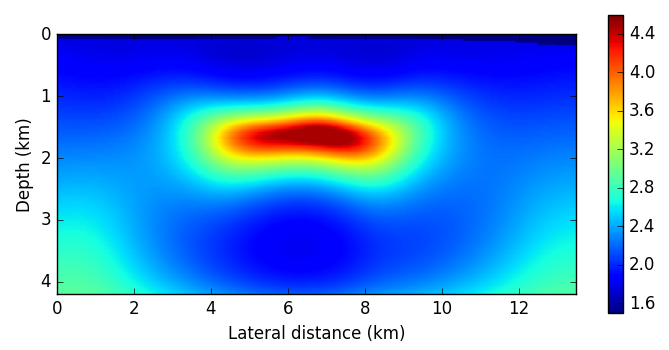}\label{fig:mfwiI}}
    \subfigure[\footnotesize The final model after stage II of the joint inversion.]{\image{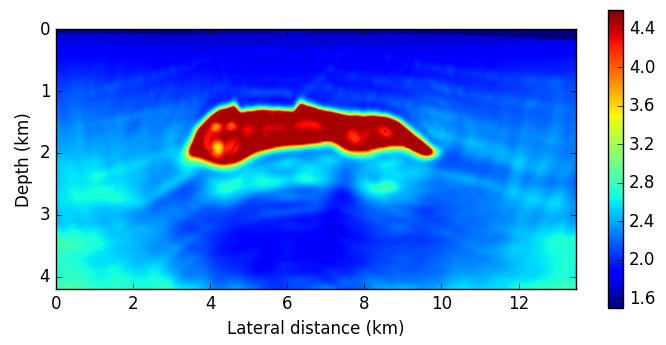}\label{fig:mfwiII}}
\end{center}
\caption{\footnotesize Recovery of the two dimensional SEG/EAGE salt body model. Velocity units are in $km/sec$.}
\label{fig:SEG_res}
\end{figure}

We apply Algorithm \ref{alg:twoPhaseJointInv} to fit the model to the data. Throughout the recovery process, in each Gauss Newton iteration we apply five preconditioned CG steps for the GN direction problem (the preconditioning is applied with the regularization Hessian matrix, in \eqref{eq:reg1} or \eqref{eq:reg2}, \tred{and the low number of CG steps also acts as regularization}). In stage $I$ we use the travel time data and the first frequency and apply 15 Gauss Newton iterations. In stage $II$, we treat the frequencies in batches of three consecutive frequencies \tred{(or travel time and two frequencies)} at the most, applying 5 Gauss Newton iterations for each batch. We apply three continuation sweeps over all of the frequencies to get the final reconstruction. \tred{The parameter $\beta$---the proportion between the two misfit terms in \eqref{eq:minjoint}---should initially be chosen large such that the tomography misfit is more dominant than the FWI misfit. Then it can be reduced as the model is improved during the minimization process to shift from the rather stable travel time tomography to the highly non-linear FWI. In this experiment we choose $\beta$ to be 2500 for the first stage, and then for stage $II$ we reduce it to 50 (other similar strategies may work as well).} We choose the regularization parameter $\alpha$ to be large ($10^4$), and decrease it by a factor of 10 between each sweep. \tred{In addition, because we apply a rather small number of CG iterations preconditioned by $\nabla^2R$, the Gauss Newton correction $\delta\bfm$ is smooth because it belongs to a smoothed Krylov subspace. As long as the number of CG iterations is small (lower than about 10 in our experiments), this plays a role in regularization.} Lower and upper bounds of 1.5 and 4.5 $km/sec$ are used for the velocity model.

\begin{figure}
\begin{center}
	\newcommand{\image}[1]{\includegraphics[width=1.0\linewidth]{#1}}
    \image{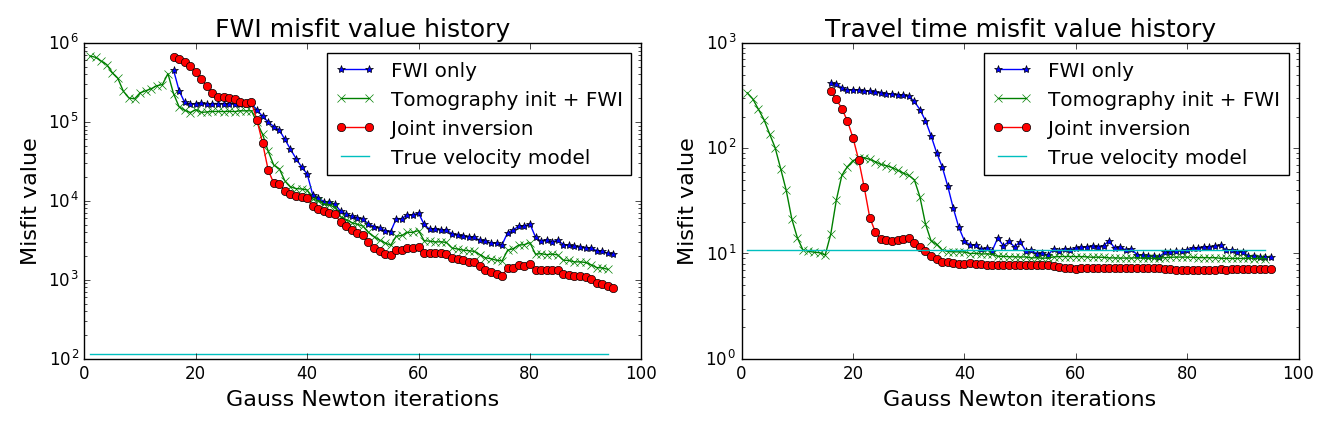}
\end{center}
\caption{\footnotesize Data fit convergence history, with respect to the FWI data (left) and travel time data (right). The FWI misfit value is calculated for all the frequencies, even though the GN iterations include only part of them.}
\label{fig:misfit}
\end{figure}

Our first experiment involves applying FWI starting from the linear model in Fig. \ref{fig:2D-2}. The reconstruction appears in Fig. \ref{fig:mFWIonly}, and includes only a thin layer of the upper part of the salt body. In our second experiment we use the travel time tomography to initialize FWI. Fig. \ref{fig:mTomoI} shows the travel time tomography result starting from the reference model in Fig. \eqref{fig:2D-2}. The inversion determines the area beneath the salt body by regularization only, and produces a salt-flooding feature in the model. The following FWI inversion cannot reasonably recover that area, leaving ``left-overs'' from the initialization, as observed in Fig. \ref{fig:mfwiTomoII}. Figure \ref{fig:mfwiI} shows the smooth model obtained by the first stage using the joint inversion starting from the linear model in Fig. \eqref{fig:2D-2}. The reconstruction is indeed a smooth version of the true model, and is considerably different than the travel time tomography reconstruction. Particularly, the joint process is able to approximately determine the low velocity underneath the salt body. The reconstructed model in Fig. \ref{fig:mfwiII} shows a good estimation of the shape of the salt body, reconstructs the area above and beneath the salt quite well. \tred{Fig. \ref{fig:misfit} shows the convergence history of the FWI and travel time data misfit values for the GN iterations of all the experiments. The FWI misfit values include all of the data even though each of the GN iterations include only part of the data in the frequency continuation procedure. In the history of the travel time misfit values, most GN iterations do not include the travel time data. The first 15 iterations of the tomography initialized FWI are the generation of the model in Fig. \ref{fig:mTomoI} starting from the model in \ref{fig:2D-2}, and these iterations appear only for this setting. The other inversions start at the model in \ref{fig:2D-2}, and their first iteration is labeled as 16.} In agreement with the visual quality of the reconstructions, the FWI-only experiment produces the highest misfit values, the joint inversion produces the lowest misfit values, and the tomography-initialized FWI produces a misfit values in between them. \tred{This corresponds to both FWI and travel time misfit values. Interestingly, the tomography initialized FWI curve shows that the FWI process initially ruined the travel time data fit, and then restored it through the FWI data fitting process. The joint inversion process produced monotonically deceasing misfit values with respect to both data sets. In addition, all the inversions ended with a model that has a misfit value that is lower than that of the true model with respect to the travel time data. This suggests that the true model is not optimal for the travel time misfit because the data has a relatively high noise level as shown in Fig. \ref{fig:travelPick}. Finally, notice that the magnitude of the FWI misfit is 2-3 orders of magnitude higher than the travel time misfit. This explains our choice of the balancing parameter $\beta$ in Eq. \eqref{eq:minjoint} to be so high, approximately in this magnitude.}

\subsection{Joint inversion in three dimensions}

\begin{figure}
\begin{center}
	\newcommand{\image}[1]{\includegraphics[width=0.48\linewidth]{#1}}
  \subfigure[\footnotesize The SEG/EAGE velocity model.]{\image{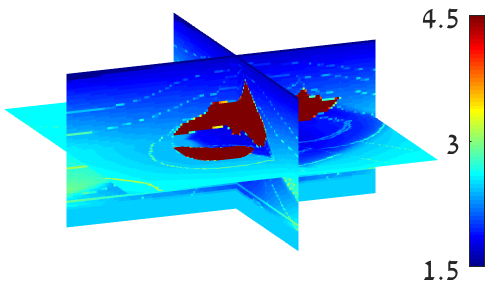}\label{fig:mtrue3D}}
  \subfigure[\footnotesize The starting linear model.]{\image{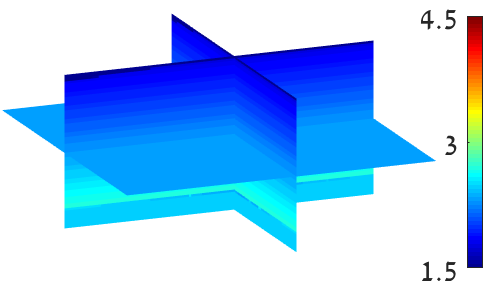}\label{fig:mref3D}}
    \subfigure[\footnotesize Smooth model obtained after stage I]{\image{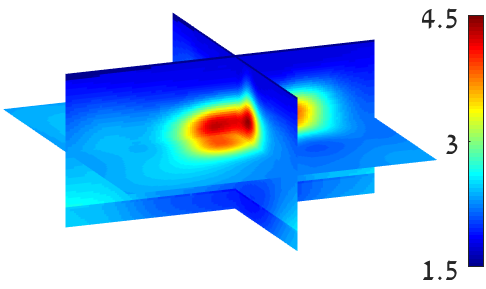}\label{fig:mfwi3DI}}
    \subfigure[\footnotesize Final model obtained after stage II]{\image{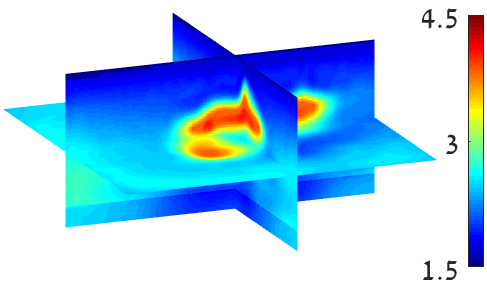}\label{fig:mfwi3DII}}
\end{center}
\caption{\footnotesize Results for 3D joint inversion. Velocity units are in $km/sec$.}
\label{fig:SEG3D}
\end{figure}

We now consider the three dimensional version of the experiment described above, using the 3D SEG/EAGE model~\cite{aminzadeh19973} as the true model for the seismic wave velocity of the subsurface, as presented in Fig. \ref{fig:mtrue3D}. The domain size is $13.5\,{\rm km}\times13.5\,{\rm km}\times4.2\,{\rm km}$, divided into $145\times145\times70$ equally-sized mesh cells $93\,{\rm m}\times93\,{\rm m}\times60\,{\rm m}$ each. To populate most of the absorbing boundary layer we add a 10-point padding to the model, resulting in a $165\times165\times80$ grid for both the forward and inverse problems. We use data corresponding to the frequencies $f_i = \{1.5,2.0\}$ Hz ($\omega_i = 2\pi f_i$). For $\bfgamma$ we again use attenuation of $0.01\cdot4\pi$, and use 14 grid points for the absorbing layer itself (10 of which lie in the model padding). We use 81 sources, arranged on a $9 \times 9$ grid located at the center of the free surface, with a distance of $1.488\,{\rm km}$ between each source. The waveform data are given at an array of $129 \times 129$ receivers located at the center of the free surface, placed $93\,{\rm m}$ apart. The data for each source are recorded only in receivers distanced 200 m to 16 km away from it. In this experiment the \tred{waveform and traveltime data are synthetically simulated using the Helmholtz and eikonal solvers (respectively)} on the same mesh for each frequency $f_i$, with 1\% Gaussian white noise added.

For the inversion, we again apply the two-stage Algorithm \ref{alg:twoPhaseJointInv}. The inversion is performed on a single machine, using the direct method MUMPS to solve the FWI forward problems \eqref{eq:helm} using a shared memory parallelism on a single worker, and multiple workers to solve the tomography forward problems \eqref{eq:factoredeikonal}. In the first stage, when we treat the travel time data together with the first frequency, we apply 10 GN iterations with 9 projected PCG iterations in each. We use the same regularization strategy as in the two dimensional case. We note that preconditioning the Hessian requires solving one or two Poisson equations at each CG iteration (two in the case of \eqref{eq:reg1}, one in the case of \eqref{eq:reg1}). This may be a non-trivial in three dimensions, both in terms of memory and computations. In this experiment we use the Smoothed Aggregation algebraic multigrid algorithm reported in \cite{treister2015non} for this task, which is also available in the multigrid package mentioned above.

Because of memory considerations, we apply the second stage of Algorithm \ref{alg:twoPhaseJointInv} in batches of two consecutive frequencies at the time (or travel time data and one frequency). We apply three continuation sweeps, using 5 GN iterations and 5 CG iterations for each inversion. \tred{The parameters $\alpha$ and $\beta$ have the same values as in the 2D experiments.} When handling each frequency, the worker solves the Helmholtz systems for all the sources in batches of only 27 sources at the time, using the same factorization, to further reduce the memory footprint (the fields, which are necessary for the sensitivities, are saved to the disk). Again we use the same regularization strategy as in 2D, using \eqref{eq:reg2}.

The joint inversion reconstruction is given in Fig.~\ref{fig:mfwi3DII}. It shows a blurred version of the true model. Including data that corresponds to higher frequencies will enable a sharper reconstruction, but will also require a finer mesh, which is much more expensive to process. \tred{To get such a reconstruction with FWI alone, one requires frequencies starting from as low as 0.5Hz \cite{jInv16}.} This inversion took about three days of computations using the shared memory computer described earlier.

\subsection{Solving the 3D Helmholtz equation for multiple right hand sides}

As discussed in the introduction, the use of frequency domain formulation is possible only when a robust solver for the Helmholtz equation is available.
In this section we demonstrate our iterative solution of the Helmholtz equation and show how the solver can benefit from having multiple right hand sides. To this end, we use the same workstation as before, and measure the time it takes to generate the data that corresponds to the model in Figure \ref{fig:mtrue3D} for multiple sources and a high frequency. The frequency is chosen such that $\frac{1}{v_{min}}\omega h_{min} = 2\pi/10$ (ten grid-points per wavelength for the lowest velocity in the model). We use an attenuation parameter of $0.02\pi$. We compare the run time of the direct solver MUMPS and three variants of the shifted Laplacian iterative multigrid solver using 3 levels (we measure the setup and average solve time). To define the shifted Helmholtz operator, we add $0.2\omega$ to $\bfgamma$ in \eqref{eq:helm}. We use 2 pre and post weighted Jacobi relaxations with a weight of 0.8. The iterative methods are run until the initial residual is reduced by $10^6$. We denote the first multigrid configuration by \emph{W-BiCG} (W-cycles preconditioning with block BiCGSTAB \cite{el2003block} as a Krylov method.) For the second and third configurations, we use the block FGMRES(5) method \cite{calandra2012flexible} preconditioned by either W-cycles or K-cycles \cite{notay2008recursive}, and denote them by \emph{W-FGMRES} and \emph{K-FGMRES} respectively. In the latter configuration the coarse grid block system is solved by two iterations of block FGMRES, a method related to the one used in \cite{calandra2013improved}. We note that some approaches use GMRES as a relaxation \cite{cools2014new}, but we found this to be quite expensive in our parallel setting compared to the Jacobi relaxations, and do not include these results.

\begin{table}
\centering
\begin{tabular}{|c|c|cc|cc|cc|}
  \hline
  \mc{8}{|c|}{Grid: {$129\times129\times65$}, Number of sources: 64}\\
  \hline
   $\#$ & {\emph{MUMPS}}& \mc{2}{|c}{\emph{W-BiCG}}& \mc{2}{c}{\emph{W-FGMRES}} &\mc{2}{c|}{\emph{K-FGMRES}}\\
   rhs  &  $t_{fact}$: 123$s$ & \mc{6}{|c|}{{ $t_{setup}$:} 14.9$s$}\\

  \hline
  & $t_{sol}$ & $\#Cyc$  & $t_{sol}$  & $\#Cyc$ & $t_{sol}$ & $\#Cyc$ & $t_{sol}$\\
 1 & 2.73s  & 25.7  & 5.44s & 23.4  & 6.16s & 22.7 & 6.25s\\
 4 & 0.97 & 26.3  & 3.47s & 23.2 &  4.4s &  22.2 &4.82s\\
 16 & 0.43 &25.7  & 2.28s & 22.2 &  3.0s &  21.5 &3.70s\\
 32 & 0.44  &  26.0  & 2.02s & 21.0 & 2.82s  &  21.0 &3.20s\\
 64 &  0.39 & 23.0  & 1.79s & 20.0 & 2.89s  &  20.0 &3.44s\\
             \hline
             \hline
  \mc{8}{|c|}{Grid: {$193\times193\times97$}, Number of sources: 64}\\
  \hline
   $\#$ & {\emph{MUMPS}}& \mc{2}{|c}{\emph{W-BiCG}}& \mc{2}{c}{\emph{W-FGMRES}} &\mc{2}{c|}{\emph{K-FGMRES}}\\
   rhs  &  $t_{fact}$: 914$s$ & \mc{6}{|c|}{{ $t_{setup}$:} 50.9$s$}\\
  \hline
   & $t_{sol}$ & $\#Cyc$  & $t_{sol}$  & $\#Cyc$ & $t_{sol}$ & $\#Cyc$ & $t_{sol}$\\
 1 & 12.2s  & 41.7  & 26.8s & 38.3  & 26.8s & 34.7 &25.5s\\
 4 & 4.42s & 43.0  & 14.5s & 38.0 &  18.2s &  34.5 &20.0s\\
 16 & 1.66s &42.0  & 11.1s & 36.7 &  15.4s &  34.0 &16.7s\\
 32 & 1.68s  & 40.5 & 9.5s & 35.5 & 14.5s  &  33.0 &15.8s\\
 64 & 1.57s  & 38.0  & 9.3s & 34.0 & 15.9s  &  32.0 &19.5s\\
  \hline
  \hline
  \mc{8}{|c|}{Grid: {$289\times289\times145$}, Number of sources: 32}\\
  \hline
   $\#$ & {\emph{MUMPS}}& \mc{2}{|c}{\emph{W-BiCG}}& \mc{2}{c}{\emph{W-FGMRES}} &\mc{2}{c|}{\emph{K-FGMRES}}\\
   rhs  &  $t_{fact}$:  & \mc{6}{|c|}{{ $t_{setup}$:} 149$s$}\\
  \hline
   & $t_{sol}$ & $\#Cyc$  & $t_{sol}$  & $\#Cyc$ & $t_{sol}$ & $\#Cyc$ & $t_{sol}$\\
 1 & --  & 74.9  & 130.5s & 68.75  & 131.5s & 58.6 &121.5s\\
 4 & -- &  76.8 & 79.5s & 68.25 &  101.9s &  58.2 &102.4s\\
 16 & -- & 74.5 & 60.7s & 68.0 &  91.0s &  58.5 & 91.3s\\
 32 & --  & 73.0 & 53.7s & -- & --  &  -- &--\\
  \hline
  \hline
  \mc{8}{|c|}{Grid: {$433\times433\times217$}, Number of sources: 16}\\
  \hline
   $\#$ & {\emph{MUMPS}}& \mc{2}{|c}{\emph{W-BiCG}}& \mc{2}{c}{\emph{W-FGMRES}} &\mc{2}{c|}{\emph{K-FGMRES}}\\
   rhs  &  $t_{fact}$:  & \mc{6}{|c|}{{ $t_{setup}$:} 512$s$}\\
  \hline
   & $t_{sol}$ & $\#Cyc$  & $t_{sol}$  & $\#Cyc$ & $t_{sol}$ & $\#Cyc$ & $t_{sol}$\\
 1 & $-$ & 136 & 744s & 128 & 816s& 101 &674s\\
 4 & $-$ & 142 & 494s  &125  & 744s&99 &622s\\
  \hline
  \hline
\end{tabular}
\label{tab:MGperformance}
\caption{The performance of the MUMPS solver and three shifted Laplacian multigrid configurations for solving multiple right hand sides.
$t_{sol}$ is the average solution time for a single right-hand-side. This time does not include the time for the setup or the factorization. $\#$ rhs denotes the number of right hand sides which are solved together in a block. The frequency for the experiments grows with the problem size.}
\end{table}

Table \ref{tab:MGperformance} summarizes our experiments with the MG solver, where the cells with `$-$' in the tables denote cases where our machine did not have enough memory to solve the block systems. The table shows an advantage in performance when solving the Helmholtz equation for an increasing number of right hand sides, whether directly or iteratively. When compared with the MUMPS solver, the iterative solution time is about five times slower, but the setup time of MUMPS is much higher. Choosing between the two solution options---iterative vs. direct---depends on whether the direct solver \tred{is feasible using the available memory}, and in the number of linear systems to be solved. A GN algorithm would generally favor a direct solution if the factorization can be stored in the memory. However, in large three dimensional problems the factorization usually cannot fit in memory and iterative methods must be used. We note that for lower frequencies, the running time of MUMPS (setup and solve) generally remains similar to that of high frequencies, but the solve time of the MG solver is much smaller depending on the frequency.

\begin{figure}
\begin{center}
	\newcommand{\image}[1]{\includegraphics[width=0.75\linewidth]{#1}}
    \image{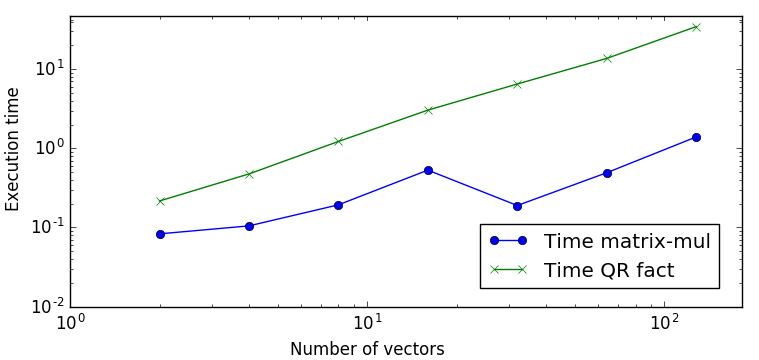}
\end{center}
\caption{\footnotesize Execution time of ``tall and thin'' matrix multiplication and QR factorization. The matrices $U,V$ are of size $n\times m$ where $n=2^{24}$ and $m$ varies between 2 to 128. Matrix multiplication ($U^TV$) is performed using {\tt BLAS.gemm!}, and the QR factorization is performed using {\tt qrfact!}. This code (and the timing) contains no memory allocation in Julia. }
\label{fig:QR}
\end{figure}

For all the iterative MG solvers, increasing the number of right-hand sides generally achieves a speed-up factor of about 2-3 when compared with the parallelized iterative solution of a single right hand side. One exception is the performance of the FGMRES configurations with 64 right-hand-sides (discussed later). Most of the improvement in the solution time is achieved by having better parallelization performance from the multi-core machine. When using a rather large number of right hand sides, the block Krylov method is able to reduce the number of iterations needed for convergence, but requires more work in the block inner products. The fastest configuration of the three options is \emph{W-BiCG} in most cases where multiple RHSs are solved simultaneously. The number of preconditioning cycles is similar in both W-cycle preconditioned options. The K-cycle preconditioning leads to less iterations, but the K-cycle requires more computational time than the W-cycle because of the block FGMRES algorithm for the coarse grid. The advantage of block BiCGSTAB is that it requires less block multiplications than block FGMRES(5) and does not require a QR factorization at each iteration. The QR factorization is much slower than a block multiplication and is the reason why the performance of FGMRES degrades for 64 RHSs. Figure \ref{fig:QR} shows the runtime comparison between a ``tall and thin'' matrix multiplication and ``thin'' QR factorization. In Julia, the QR factorization is at least one order of magnitude more expensive than a matrix multiplication, which explains the increased timings of block FGMRES compared to BiCGSTAB (which does not perform QR). Similar timings are also obtained using MATLAB.

One shortcoming of BiCGSTAB compared to FGMRES is that it is not flexible, and its preconditioner must remain stationary. This requires the relaxation method and coarsest grid solution to be stationary. Because matrix factorizations are expensive in memory (also for the coarsest grid), the works \cite{calandra2012flexible,calandra2013improved} suggest an iterative Jacobi-preconditioned GMRES solver for the coarsest grid. Even though this method is not optimal, the overall performance of the MG solver is acceptable because the coarsest matrix is relatively small and performing many iterations is not expensive. In this regard, part of our future effort will focus on a stationary method for efficiently solving the shifted coarsest grid system, which in principle is easier to solve than the original Helmholtz system.

\section{Conclusions and further work}
\label{sec6}
In this work we explore a methodology that aids full waveform inversion to converge to a plausible minimum in the absence of low frequency data. The method is based on (1) using travel time data instead of the missing low frequency waveform data in a joint inversion of FWI and travel time tomography, and (2) using a fourth order regularized inversion in a preliminary stage to obtain the low-frequency model. Our experiments show that solving the joint inversion problem can lead to a better recovery of the underlying medium compared to recoveries obtained by solving FWI after initialization with a linear model or a travel time tomogram. Additionally, since we jointly fit the full waveform and travel time data, our final model is consistent
for both physical models. In 3D, the added computational effort that is required to solve the joint inversion is negligible compared to that required to solve FWI alone.

Our method has two main limitations: (1) having a meaningful travel time tomography inversion requires the recording of relatively long offset data and travel time picking. While picking in marine data is relatively simple, it can be a more involved process for land based data. (2) While our preliminary stage is robust, the optimization in the second stage may still reach a local minimum because the highly non-linear FWI problem is the dominant problem in this stage. Our future research aims at relieving some of this non-linearity and making the process more stable.

Lastly, we considered the main computational bottleneck in FWI---the solution of the Helmholtz forward problems, \tred{and compared the available strategies} to exploit the multiple right hand sides that need to be solved in FWI for accelerating the iterative solution of the Helmholtz linear systems. Our experiments show that the best performance gain is obtained by using block BiCGSTAB with a stationary MG cycle. In this regard, our future research aims to improve the scalability of the solver, and develop an efficient and parallel stationary inexact solver to the coarsest grid.

\section{Acknowledgements}
The authors would like to thank Prof. Mauricio Sacchi, University of Alberta, and Dr. Elliot Holtham, UBC, for their help and guidance throughout this work. The research leading to these results received funding from the European Union's - Seventh Framework Programme (FP7 / 2007-2013) under grant agreement no 623212 - MC Multiscale Inversion. 

\bibliographystyle{siam}
\bibliography{JointV3}

\begin{thebibliography}{10}

\bibitem{ricker}
{\em {Ricker Wavelet}}.
\newblock \url{http://wiki.seg.org/wiki/Dictionary:Ricker_wavelet/}, 2013.
\newblock [Online; accessed 21-March-2017].

\bibitem{MUMPS2001}
{\sc P.~R. Amestoy, I.~S. Duff, J.-Y. L'Excellent, and J.~Koster}, {\em A fully
  asynchronous multifrontal solver using distributed dynamic scheduling}, SIAM
  Journal on Matrix Analysis and Applications, 23 (2001), pp.~15--41.

\bibitem{aminzadeh19973}
{\sc F.~Aminzadeh, B.~Jean, and T.~Kunz}, {\em 3-D salt and overthrust models},
  Society of Exploration Geophysicists, 1997.

\bibitem{baker2006improving}
{\sc A.~H. Baker, J.~M. Dennis, and E.~R. Jessup}, {\em On improving linear
  solver performance: A block variant of {GMRES}}, SIAM Journal on Scientific
  Computing, 27 (2006), pp.~1608--1626.

\bibitem{benaichouche2015first}
{\sc A.~Benaichouche, M.~Noble, and A.~Gesret}, {\em First arrival traveltime
  tomography using the fast marching method and the adjoint state technique},
  in 77th EAGE Conference Proceedings., 2015.

\bibitem{biondi2014simultaneous}
{\sc B.~Biondi and A.~Almomin}, {\em Simultaneous inversion of full data
  bandwidth by tomographic full-waveform inversion}, Geophysics, 79 (2014),
  pp.~WA129--WA140.

\bibitem{boonyasiriwat2010applications}
{\sc C.~Boonyasiriwat, G.~T. Schuster, P.~Valasek, and W.~Cao}, {\em
  Applications of multiscale waveform inversion to marine data using a flooding
  technique and dynamic early-arrival windows}, Geophysics, 75 (2010),
  pp.~R129--R136.

\bibitem{BHM00}
{\sc W.~L. Briggs, V.~E. Henson, and S.~F. McCormick}, {\em A multigrid
  tutorial}, SIAM, second~ed., 2000.

\bibitem{calandra2012flexible}
{\sc H.~Calandra, S.~Gratton, J.~Langou, X.~Pinel, and X.~Vasseur}, {\em
  Flexible variants of block restarted {GMRES} methods with application to
  geophysics}, SIAM Journal on Scientific Computing, 34 (2012), pp.~A714--A736.

\bibitem{calandra2013improved}
{\sc H.~Calandra, S.~Gratton, X.~Pinel, and X.~Vasseur}, {\em An improved
  two-grid preconditioner for the solution of three-dimensional {Helmholtz}
  problems in heterogeneous media}, Numerical Linear Algebra with Applications,
  20 (2013), pp.~663--688.

\bibitem{cools2014new}
{\sc S.~Cools, B.~Reps, and W.~Vanroose}, {\em A new level-dependent coarse
  grid correction scheme for indefinite {Helmholtz} problems}, Numerical Linear
  Algebra with Applications, 21 (2014), pp.~513--533.

\bibitem{el2003block}
{\sc A.~El~Guennouni, K.~Jbilou, and H.~Sadok}, {\em A block version of
  {BiCGSTAB} for linear systems with multiple right-hand sides}, Electronic
  Transactions on Numerical Analysis, 16 (2003), pp.~129--142.

\bibitem{EpanomeritakisAkcelikGhattasBielak2008}
{\sc I.~Epanomeritakis, V.~Akcelik, O.~Ghattas, and J.~Bielak}, {\em A
  {Newton-CG} method for large-scale three-dimensional elastic full-waveform
  seismic inversion}, Inverse Problems,  (2008).

\bibitem{erlangga2006novel}
{\sc Y.~A. Erlangga, C.~W. Oosterlee, and C.~Vuik}, {\em A novel multigrid
  based preconditioner for heterogeneous {Helmholtz} problems}, SIAM Journal on
  Scientific Computing, 27 (2006), pp.~1471--1492.

\bibitem{fomel2009fast}
{\sc S.~Fomel, S.~Luo, and H.~Zhao}, {\em Fast sweeping method for the factored
  eikonal equation}, Journal of Computational Physics, 228 (2009),
  pp.~6440--6455.

\bibitem{haber2014computational}
{\sc E.~Haber}, {\em Computational Methods in Geophysical Electromagnetics},
  vol.~1, SIAM, 2014.

\bibitem{hao2000}
{\sc E.~Haber, U.~Ascher, and D.~Oldenburg}, {\em On optimization techniques
  for solving nonlinear inverse problems}, Inverse problems, 16 (2000),
  pp.~1263--1280.

\bibitem{haber2011fast}
{\sc E.~Haber and S.~MacLachlan}, {\em A fast method for the solution of the
  {Helmholtz} equation}, Journal of Computational Physics, 230 (2011),
  pp.~4403--4418.

\bibitem{knibbe2011gpu}
{\sc H.~Knibbe, C.~W. Oosterlee, and C.~Vuik}, {\em {GPU} implementation of a
  {Helmholtz} {Krylov} solver preconditioned by a shifted laplace multigrid
  method}, Journal of Computational and Applied Mathematics, 236 (2011),
  pp.~281--293.

\bibitem{krebs09ffw}
{\sc J.~R. Krebs, J.~E. Anderson, D.~Hinkley, R.~Neelamani, S.~Lee,
  A.~Baumstein, and M.-D. Lacasse}, {\em Fast full-wavefield seismic inversion
  using encoded sources}, Geophysics, 74 (2009), pp.~WCC177--WCC188.

\bibitem{lambare2008stereotomography}
{\sc G.~Lambar{\'e}}, {\em Stereotomography}, Geophysics, 73 (2008),
  pp.~VE25--VE34.

\bibitem{lambare2004stereotomography}
{\sc G.~Lambar{\'e}, M.~Alerini, R.~Baina, and P.~Podvin}, {\em
  Stereotomography: a semi-automatic approach for velocity macromodel
  estimation}, Geophysical Prospecting, 52 (2004), pp.~671--681.

\bibitem{li2013first}
{\sc S.~Li, A.~Vladimirsky, and S.~Fomel}, {\em First-break traveltime
  tomography with the double-square-root eikonal equation}, Geophysics, 78
  (2013), pp.~U89--U101.

\bibitem{liu2015joint}
{\sc Z.~Liu and J.~Zhang}, {\em Joint traveltime and waveform envelope
  inversion for near-surface imaging}, in 2015 SEG Annual Meeting, Society of
  Exploration Geophysicists, 2015.

\bibitem{luo2011factored}
{\sc S.~Luo and J.~Qian}, {\em Factored singularities and high-order
  lax--friedrichs sweeping schemes for point-source traveltimes and
  amplitudes}, Journal of Computational Physics, 230 (2011), pp.~4742--4755.

\bibitem{luo2012fast}
\leavevmode\vrule height 2pt depth -1.6pt width 23pt, {\em Fast sweeping
  methods for factored anisotropic eikonal equations: multiplicative and
  additive factors}, Journal of Scientific Computing, 52 (2012), pp.~360--382.

\bibitem{LouQianBurridge2014}
{\sc S.~Luo, J.~Qian, and R.~Burridge}, {\em High-order factorization based
  high-order hybrid fast sweeping methods for point-source eikonal equations},
  SIAM Journal on Numerical Analysis, 52 (2014), pp.~23--44.

\bibitem{metivier2016measuring}
{\sc L.~M{\'e}tivier, R.~Brossier, Q.~M{\'e}rigot, E.~Oudet, and J.~Virieux},
  {\em Measuring the misfit between seismograms using an optimal transport
  distance: application to full waveform inversion}, Geophysical Journal
  International, 205 (2016), pp.~345--377.

\bibitem{nw}
{\sc J.~Nocedal and S.~Wright}, {\em Numerical Optimization}, Springer, New
  York, 1999.

\bibitem{notay2008recursive}
{\sc Y.~Notay and P.~S. Vassilevski}, {\em Recursive {Krylov}-based multigrid
  cycles}, Numerical Linear Algebra with Applications, 15 (2008), pp.~473--487.

\bibitem{oosterlee2010shifted}
{\sc C.~Oosterlee, C.~Vuik, W.~Mulder, and R.-E. Plessix}, {\em
  Shifted-laplacian preconditioners for heterogeneous {Helmholtz} problems}, in
  Advanced Computational Methods in Science and Engineering, Springer, 2010,
  pp.~21--46.

\bibitem{operto20073D}
{\sc S.~Operto, J.~Virieux, P.~Amestoy, J.-Y. L’Excellent, L.~Giraud, and
  H.~B.~H. Ali}, {\em 3d finite-difference frequency-domain modeling of
  visco-acoustic wave propagation using a massively parallel direct solver: A
  feasibility study}, Geophysics, 72 (2007), pp.~SM195--SM211.

\bibitem{poulson2013parallel}
{\sc J.~Poulson, B.~Engquist, S.~Li, and L.~Ying}, {\em A parallel sweeping
  preconditioner for heterogeneous {3D} {Helmholtz} equations}, SIAM Journal on
  Scientific Computing, 35 (2013), pp.~C194--C212.

\bibitem{pratt1999}
{\sc R.~Pratt}, {\em Seismic waveform inversion in the frequency domain, part
  1: Theory, and verification in a physical scale model}, Geophysics, 64
  (1999), pp.~888--901.

\bibitem{pratt1998gauss}
{\sc R.~G. Pratt, C.~Shin, and G.~Hick}, {\em {Gauss--Newton} and full {Newton}
  methods in frequency--space seismic waveform inversion}, Geophysical Journal
  International, 133 (1998), pp.~341--362.

\bibitem{Tobias2016}
{\sc B.~Reps and T.~Weinzierl}, {\em Complex additive geometric multilevel
  solvers for {Helmholtz} equations on spacetrees},  (2016).

\bibitem{rudin1992nonlinear}
{\sc L.~I. Rudin, S.~Osher, and E.~Fatemi}, {\em Nonlinear total variation
  based noise removal algorithms}, Physica D: Nonlinear Phenomena, 60 (1992),
  pp.~259--268.

\bibitem{jInv16}
{\sc L.~Ruthotto, E.~Treister, and E.~Haber}, {\em {jInv} -- a flexible {Julia}
  package for {PDE} parameter estimation}, Submitted,  (2016).

\bibitem{saad1993flexible}
{\sc Y.~Saad}, {\em A flexible inner-outer preconditioned {GMRES} algorithm},
  SIAM Journal on Scientific Computing, 14 (1993), pp.~461--469.

\bibitem{sabbione2010automatic}
{\sc J.~I. Sabbione and D.~Velis}, {\em Automatic first-breaks picking: New
  strategies and algorithms}, Geophysics, 75 (2010), pp.~V67--V76.

\bibitem{saragiotis2013automatic}
{\sc C.~Saragiotis, T.~Alkhalifah, and S.~Fomel}, {\em Automatic traveltime
  picking using instantaneous traveltime}, Geophysics, 78 (2013), pp.~T53--T58.

\bibitem{schenk2004solving}
{\sc O.~Schenk and K.~G{\"a}rtner}, {\em Solving unsymmetric sparse systems of
  linear equations with pardiso}, Future Generation Computer Systems, 20
  (2004), pp.~475--487.

\bibitem{sethian1996fast}
{\sc J.~A. Sethian}, {\em A fast marching level set method for monotonically
  advancing fronts}, Proceedings of the National Academy of Sciences, 93
  (1996), pp.~1591--1595.

\bibitem{sethian1999fast}
\leavevmode\vrule height 2pt depth -1.6pt width 23pt, {\em Fast marching
  methods}, SIAM review, 41 (1999), pp.~199--235.

\bibitem{simoncini1995iterative}
{\sc V.~Simoncini and E.~Gallopoulos}, {\em An iterative method for
  nonsymmetric systems with multiple right-hand sides}, SIAM Journal on
  Scientific Computing, 16 (1995), pp.~917--933.

\bibitem{somersalo2004statistical}
{\sc E.~Somersalo and J.~Kaipio}, {\em Statistical and computational inverse
  problems}, Applied Mathematical Sciences, 160 (2004).

\bibitem{taran}
{\sc A.~Tarantola}, {\em Inverse problem theory}, Elsevier, Amsterdam, 1987.

\bibitem{TH2016}
{\sc E.~Treister and E.~Haber}, {\em A fast marching algorithm for the factored
  eikonal equation}, Accepted to Journal of Computational Physics,  (2016).

\bibitem{treister2015non}
{\sc E.~Treister and I.~Yavneh}, {\em Non-galerkin multigrid based on
  sparsified smoothed aggregation}, SIAM Journal on Scientific Computing, 37
  (2015), pp.~A30--A54.

\bibitem{TOS01}
{\sc U.~Trottenberg, C.~Oosterlee, and A.~Sch\"{u}ller}, {\em Multigrid},
  Academic Press, London and San Diego, 2001.

\bibitem{van1992bi}
{\sc H.~A. Van~der Vorst}, {\em {Bi-CGSTAB}: A fast and smoothly converging
  variant of {Bi-CG} for the solution of nonsymmetric linear systems}, SIAM
  Journal on scientific and Statistical Computing, 13 (1992), pp.~631--644.

\bibitem{van2013mitigating}
{\sc T.~van Leeuwen and F.~J. Herrmann}, {\em Mitigating local minima in
  full-waveform inversion by expanding the search space}, Geophysical Journal
  International, 195 (2013), pp.~661--667.

\bibitem{van20143d}
\leavevmode\vrule height 2pt depth -1.6pt width 23pt, {\em {3D}
  frequency-domain seismic inversion with controlled sloppiness}, SIAM Journal
  on Scientific Computing, 36 (2014), pp.~S192--S217.

\bibitem{van2016}
{\sc T.~van Leeuwen and F.~J. Herrmann}, {\em A penalty method for
  {PDE}-constrained optimization in inverse problems}, Inverse Problems, 32
  (2016), p.~015007.

\bibitem{virieux2009overview}
{\sc J.~Virieux and S.~Operto}, {\em An overview of full-waveform inversion in
  exploration geophysics}, Geophysics, 74 (2009), pp.~WCC1--WCC26.

\bibitem{vogel2002computational}
{\sc C.~R. Vogel}, {\em Computational methods for inverse problems}, vol.~23,
  SIAM, Philadelphia, 2002.

\bibitem{WahbaBook}
{\sc G.~Wahba}, {\em Spline Models for Observational Data}, Society for
  Industrial and Applied Mathematics, 1990.

\bibitem{wang1997sensitivities}
{\sc Y.~Wang and R.~G. Pratt}, {\em Sensitivities of seismic traveltimes and
  amplitudes in reflection tomography}, Geophysical journal international, 131
  (1997), pp.~618--642.

\bibitem{WarnerEtAt2013}
{\sc M.~Warner, A.~Ratcliffe, T.~Nangoo, J.~Morgan, A.~Umpleby, N.~Shah,
  V.~Vinje, I.~{\v{S}}tekl, L.~Guasch, C.~Win, et~al.}, {\em Anisotropic {3D}
  full-waveform inversion}, Geophysics, 78 (2013), pp.~R59--R80.

\bibitem{Yav06}
{\sc I.~Yavneh}, {\em Why multigrid methods are so efficient}, IEEE: Computing
  in Science and Engineering, 8 (2006), pp.~12--22.

\end{thebibliography}

\end{document}